\documentclass[twocolumn,twocolappendix,astrosymb]{aastex631}

\usepackage{amsmath}
\usepackage{amssymb}	% Extra maths symbols
\usepackage{natbib}
\usepackage[figuresright]{rotating}
\usepackage{threeparttable}
\usepackage{subfigure}
\usepackage{soul}

\newcommand{\s}{{s$^{-1}$}}
\newcommand{\kms}{km\,s$^{-1}$}

\newcommand{\ciii}{\ion{C}{3}]}
\newcommand{\fciii}{[\ion{C}{3}]}
\newcommand{\civ}{\ion{C}{4}}
\newcommand{\lya}{Ly$\alpha$}
\newcommand{\heii}{\ion{He}{2}}
\newcommand{\oii}{[\ion{O}{2}]}
\newcommand{\oiii}{\ion{O}{3}]}
\newcommand{\foiii}{[\ion{O}{3}]}
\newcommand{\oiiiuv}{\ion{O}{3}]}

\newcommand{\hei}{\ion{He}{1}}
\newcommand{\ha}{H$\alpha$}
\newcommand{\hb}{H$\beta$}
\newcommand{\hg}{H$\gamma$}
\newcommand{\hd}{H$\delta$}
\newcommand{\siii}{[\ion{S}{3}]}

\newcommand{\mgii}{\ion{Mg}{2}}

\newcommand{\neiii}{[\ion{Ne}{3}]}

\newcommand{\nv}{\ion{N}{5}}

\begin{document}

\title{Characterizing the Average Interstellar Medium Conditions of Galaxies at $z\sim$ 5.6 -- 9\\ with UV and Optical Nebular Lines}

\author[0000-0003-3424-3230]{Weida Hu}
\email{weidahu@tamu.edu}
\affiliation{Department of Physics and Astronomy, Texas A\&M University, College Station, TX 77843-4242, USA}
\affiliation{George P. and Cynthia Woods Mitchell Institute for Fundamental Physics and Astronomy, Texas A\&M University, College Station, TX 77843-4242, USA}

\author[0000-0003-3424-3230]{Casey Papovich}
\affiliation{Department of Physics and Astronomy, Texas A\&M University, College Station, TX 77843-4242, USA}
\affiliation{George P. and Cynthia Woods Mitchell Institute for Fundamental Physics and Astronomy, Texas A\&M University, College Station, TX 77843-4242, USA}

\author[0000-0001-5414-5131]{Mark Dickinson}
\affiliation{NSF's National Optical-Infrared Astronomy Research Laboratory, 950 N. Cherry Ave., Tucson, AZ 85719, USA}

\author[0000-0001-5448-1821]{Robert Kennicutt}
\affiliation{Department of Physics and Astronomy, Texas A\&M University, College Station, TX 77843-4242, USA}
\affiliation{George P. and Cynthia Woods Mitchell Institute for Fundamental Physics and Astronomy, Texas A\&M University, College Station, TX 77843-4242, USA}
\affiliation{Department of Astronomy and Steward Observatory, University of Arizona, Tucson, AZ 85721, USA}

\author[0000-0003-3424-3230]{Lu Shen}
\affiliation{Department of Physics and Astronomy, Texas A\&M University, College Station, TX 77843-4242, USA}
\affiliation{George P. and Cynthia Woods Mitchell Institute for Fundamental Physics and Astronomy, Texas A\&M University, College Station, TX 77843-4242, USA}

\author[0000-0001-5758-1000]{Ricardo O. Amor\'{i}n}
\affiliation{ARAID Foundation. Centro de Estudios de F\'{\i}sica del Cosmos de Arag\'{o}n (CEFCA), Unidad Asociada al CSIC, Plaza San Juan 1, E--44001 Teruel, Spain}
\affiliation{Departamento de Astronom\'{i}a, Universidad de La Serena, Av. Juan Cisternas 1200 Norte, La Serena 1720236, Chile}

\author[0000-0002-7959-8783]{Pablo Arrabal Haro}
\affiliation{NSF's National Optical-Infrared Astronomy Research Laboratory, 950 N. Cherry Ave., Tucson, AZ 85719, USA}

\author[0000-0002-9921-9218]{Micaela B. Bagley}
\affiliation{Department of Astronomy, The University of Texas at Austin, Austin, TX, USA}

\author[0000-0003-0883-2226]{Rachana Bhatawdekar}
\affiliation{European Space Agency (ESA), European Space Astronomy Centre (ESAC), Camino Bajo del Castillo s/n, 28692 Villanueva de la Cañada, Madrid, Spain}

\author[0000-0001-7151-009X]{Nikko J. Cleri}
\affiliation{Department of Physics and Astronomy, Texas A\&M University, College Station, TX 77843-4242, USA}
\affiliation{George P. and Cynthia Woods Mitchell Institute for Fundamental Physics and Astronomy, Texas A\&M University, College Station, TX 77843-4242, USA}

\author[0000-0002-6348-1900]{Justin W. Cole}
% \altaffiliation{NASA FINESST Investigator}
\affiliation{Department of Physics and Astronomy, Texas A\&M University, College Station, TX 77843-4242, USA}
\affiliation{George P. and Cynthia Woods Mitchell Institute for Fundamental Physics and Astronomy, Texas A\&M University, College Station, TX 77843-4242, USA}

\author[0000-0003-4174-0374]{Avishai Dekel}
\affiliation{Racah Institute of Physics, The Hebrew University of Jerusalem, Jerusalem 91904, Israel}

\author[0000-0002-6219-5558]{Alexander de la Vega}
\affiliation{Department of Physics and Astronomy, University of California, 900 University Ave, Riverside, CA 92521, USA}

\author[0000-0001-8519-1130]{Steven L. Finkelstein}
\affiliation{Department of Astronomy, The University of Texas at Austin, Austin, TX, USA}

\author[0000-0001-9440-8872]{Norman A. Grogin}
\affiliation{Space Telescope Science Institute, 3700 San Martin Drive, Baltimore, MD 21218, USA}

\author[0000-0001-6145-5090]{Nimish P. Hathi}
\affiliation{Space Telescope Science Institute, 3700 San Martin Drive, Baltimore, MD 21218, USA}

\author[0000-0002-3301-3321]{Michaela Hirschmann}
\affiliation{Institute of Physics, Laboratory of Galaxy Evolution, Ecole Polytechnique Fédérale de Lausanne (EPFL), Observatoire de Sauverny, 1290 Versoix, Switzerland}

\author[0000-0002-4884-6756]{Benne W. Holwerda}
\affil{Physics \& Astronomy Department, University of Louisville, 40292 KY, Louisville, USA}

\author[0000-0001-6251-4988]{Taylor A. Hutchison}
\affiliation{Astrophysics Science Division, NASA Goddard Space Flight Center, 8800 Greenbelt Rd, Greenbelt, MD 20771, USA}

\author[0000-0003-1187-4240]{Intae Jung}
\affiliation{Space Telescope Science Institute, 3700 San Martin Drive, Baltimore, MD 21218, USA}

\author[0000-0002-6610-2048]{Anton M. Koekemoer}
\affiliation{Space Telescope Science Institute, 3700 San Martin Drive, Baltimore, MD 21218, USA}

\author[0000-0001-9187-3605]{Jeyhan S. Kartaltepe}
\affiliation{Laboratory for Multiwavelength Astrophysics, School of Physics and Astronomy, Rochester Institute of Technology, 84 Lomb Memorial Drive, Rochester, NY 14623, USA}

\author[0000-0003-1581-7825]{Ray A. Lucas}
\affiliation{Space Telescope Science Institute, 3700 San Martin Drive, Baltimore, MD 21218, USA}

\author[0000-0003-1354-4296]{Mario Llerena}
\affiliation{INAF - Osservatorio Astronomico di Roma, via di Frascati 33, 00078 Monte Porzio Catone, Italy}

\author[0000-0002-9572-7813]{S. Mascia}
\affiliation{INAF - Osservatorio Astronomico di Roma, via di Frascati 33, 00078 Monte Porzio Catone, Italy}
\affiliation{Dipartimento di Fisica, Università di Roma Tor Vergata,
Via della Ricerca Scientifica, 1, 00133, Roma, Italy}

\author[0000-0001-5846-4404]{Bahram Mobasher}
\affiliation{Department of Physics and Astronomy, University of California, 900 University Ave, Riverside, CA 92521, USA}

\author[0000-0002-8951-4408]{L. Napolitano}
\affiliation{INAF - Osservatorio Astronomico di Roma, via di Frascati 33, 00078 Monte Porzio Catone, Italy}
\affiliation{Dipartimento di Fisica, Università di Roma Sapienza, Città Universitaria di Roma - Sapienza, Piazzale Aldo Moro, 2, 00185, Roma, Italy}

\author[0000-0001-8684-2222]{Jeffrey A.\ Newman}
\affiliation{Department of Physics and Astronomy and PITT PACC, University of Pittsburgh, Pittsburgh, PA 15260, USA}

\author[0000-0001-8940-6768]{Laura Pentericci}
\affiliation{INAF - Osservatorio Astronomico di Roma, via di Frascati 33, 00078 Monte Porzio Catone, Italy}

\author[0000-0003-4528-5639]{Pablo G. P\'erez-Gonz\'alez}
\affiliation{Centro de Astrobiolog\'{\i}a (CAB), CSIC-INTA, Ctra. de Ajalvir km 4, Torrej\'on de Ardoz, E-28850, Madrid, Spain}

\author[0000-0002-1410-0470]{Jonathan R. Trump}
\affiliation{Department of Physics, 196 Auditorium Road, Unit 3046, University of Connecticut, Storrs, CT 06269, USA}

\author[0000-0003-3903-6935]{Stephen M.~Wilkins} 
\affiliation{Astronomy Centre, University of Sussex, Falmer, Brighton BN1 9QH, UK}
\affiliation{Institute of Space Sciences and Astronomy, University of Malta, Msida MSD 2080, Malta}

\author[0000-0003-3466-035X]{{L. Y. Aaron} {Yung}}
\affiliation{Space Telescope Science Institute, 3700 San Martin Drive, Baltimore, MD 21218, USA}

\begin{abstract}
Ultraviolet (UV; rest-frame $\sim1200-2000$ \AA) spectra provide a wealth of diagnostics to characterize fundamental galaxy properties, such as their chemical enrichment, the nature of their stellar populations, and their amount of Lyman-continuum (LyC) radiation. 
In this work, we leverage publicly released JWST data to construct the rest-frame UV-to-optical composite spectrum of a sample of 63 galaxies at $5.6<z<9$, spanning the wavelength range from 1500 to 5200 \AA.
Based on the composite spectrum, we derive an average dust attenuation $E(B-V)_\mathrm{gas}=0.16^{+0.10}_{-0.11}$ from \hb/\hg, electron density $n_e = 570^{+510}_{-290}$ cm$^{-3}$ from the \oii\ doublet ratio, electron temperature $T_e = 17000^{+1500}_{-1500}$ K from the \foiii~$\lambda4363$/\foiii~$\lambda5007$ ratio, and an ionization parameter $\log(U)=-2.18^{+0.03}_{-0.03}$ from the \foiii/\oii\ ratio. 
Using a direct $T_e$ method, we calculate an oxygen abundance $12+\log\mathrm{(O/H)}=7.67\pm0.08$ and the carbon-to-oxygen (C/O) abundance ratio $\log\mathrm{(C/O)}=-0.87^{+0.13}_{-0.10}$.
This C/O ratio is smaller than compared to $z=0$ and $z=2$ -- 4 star-forming galaxies, albeit with moderate significance.
This indicates the reionization-era galaxies might be undergoing a rapid build-up of stellar mass with high specific star-formation rates. 
%The relatively high supernovae rate enhances oxygen, while carbon enrichment from longer-lived AGB stars has not fully proceeded.
A UV diagnostic based on the ratios of \ciii~$\lambda\lambda1907,1909$/\heii~$\lambda1640$ versus \oiii~$\lambda1666$/\heii~$\lambda1640$ suggests that the star formation is the dominant source of ionization, similar to the local extreme dwarf galaxies and $z\sim2$ -- 4 \heii--detected galaxies.
The \foiii/\oii\ and \civ/\ciii\ ratios of the composite spectrum are marginally larger than the criteria used to select galaxies as LyC leakers, 
suggesting that some of the galaxies in our sample are strong contributors to the reionizing radiation. 
\end{abstract}

\section{Introduction} \label{sec:intro}
One of the most important questions in modern astronomy is how the galaxies in the early Universe evolve and contribute to cosmic reionization. 
Those galaxies are expected to have an extreme environment capable of producing large amounts of high-energy photons and building up a highly ionized interstellar medium (ISM) that allows the escape of high-energy photons.
The highly ionized gas nebulae can also produce a number of emission lines, of which the strengths are determined by the abundance of each species \citep{Maiolino2019}, the physical condition of gas nebulae \citep{Kewley2019,Berg2021,Mingozzi2022}, dust attenuation \citep{Buat2002}, and the nature of ionizing sources \citep{Jaskot2016,Feltre2016,Xiao2018, Byler2020}, providing keys to understand the extreme environment in those reionization-era galaxies.

Before the commissioning of the James Webb Space Telescope \citep[JWST;][]{Gardner2006,Gardner2023}, early exploration of nebular emission lines in reionization-era galaxies mostly focused on the rest-frame UV lines acquired by ground-based telescopes \citep[e.g.,][]{Stark2015,Stark2017,Laporte2017,Hu2017,Mainali2017,Hutchison2019,Topping2021}.
The detections of high-ionization UV lines (\nv, \civ, \heii, \ciii) in these works indicate a lower metallicity and a higher ionization field, and hint at the possible activity of central massive black holes or the presence of Population III stars.
However, due to the high sky background and strong telluric absorption, only a small number of detections have been achieved and usually only one of the high-ionization lines is detected, leading to large uncertainty on galaxy properties.
Indirect constraints from optical emission lines have also been performed based on the mid-infrared multi-band photometry.
The observed mid-infrared colors of high-redshift galaxies reveal a significant contribution of \foiii\ $\lambda\lambda$4959,5007 and \hb\ lines \citep[e.g.,][]{Gonzalez2012,Roberts-Borsani2016,Smit2016,Bridge2019,Endsley2021}, implying a high star formation rate and a highly ionized ISM. 
However, the broadband photometry cannot resolve into individual lines, limiting a comprehensive analysis of galaxy properties.

The advent of JWST enables the detection of rest-frame UV and optical emission lines of reionization-era galaxies at an unprecedented high signal-to-noise (S/N) ratio and spectral resolution. 
With the public release of the first JWST data, the rest-frame optical emission lines have been intensively studied with a large sample of galaxies at $z\gtrsim6$ and have revealed that high-redshift galaxies exhibit generally low oxygen abundance, high ionization, high electron density, and high temperature \citep[e.g.,][]{Curti2023,Rhoads2023,Tang2023,Trump2023,Sanders2023,Cameron2023,Williams2023,Fujimoto2023,Bunker2023,Isobe2023,Jung2023}. 
In contrast, the rest-frame UV lines, providing complementary information on carbon and nitrogen abundance \citep{Feltre2016,Jones2023,Isobe2023b,Arellano-Cordova2022,Hirschmann2019,Hirschmann2023}, ionization source classification \citep{Bunker2023,Senchyna2023,Larson2023}, and ionizing photon escaping \citep{Plat2019}, but are considerably weaker compared to the strong optical lines. 
Nonetheless, we are able to recover the average UV line properties from a high S/N composite spectrum of a large sample of galaxies. 
In this work, we leverage the publicly released spectroscopic data of reionization-era galaxies from the Cosmic Evolution Early Release Science Survey\footnote{\url{https://ceers.github.io}} \citep[CEERS;][]{Finkelstein2023} and the JWST Advanced Deep Extragalactic Survey\footnote{\url{https://jades-survey.github.io/}} \citep[JADES;][]{Eisenstein2023,Bunker2023} to construct the composite spectrum and investigate the rest-frame UV to optical lines.

This paper is organized as follows.
We describe the sample in Section \ref{sec:sample}. In Section \ref{sec:stack}, we present our method to generate the composite spectrum and measure the line fluxes. 
In Section \ref{sec:results}, we present the measurements of galaxy properties, including dust attenuation, electron density and temperature, and carbon and oxygen abundances.
In Section \ref{sec:discussion}, we discuss the evolution of the carbon-to-oxygen (C/O) abundance ratio, the ionization diagnostic, and the ionizing photon leakage.
% Throughout this work, we adopt a flat $\Lambda$CDM cosmology with $\Omega_m=0.3$, $\Omega_\Lambda = 0.7$, and H$_0 = 70$ \kms\ Mpc$^{-1}$.

\section{Data and Sample} \label{sec:sample}

\subsection{NIRSpec Data from CEERS and JADES}

The CEERS and JADES NIRSpec observations utilize several combinations of dispersers and filters to achieve low ($R\sim100$; prism), medium ($R\sim1000$; G140M/F100LP, G140M/F070LP, G235M/F170LP, G395M/F290LP), and high resolutions ($R\sim2700$; G395H/F290LP, JADES only) spanning a wavelength coverage of 1 -- 5.3 $\mu$m and 0.7 -- 5.3 $\mu$m.
The goal of this work is to detect and resolve the UV nebular lines by stacking the two-dimensional (2D) NIRSpec spectra.
Therefore, we adopt the medium-resolution grating spectra (hereafter M-Grating) from these two surveys.

The CEERS M-Grating data are obtained with the G140M/F100LP, G235M/F170LP, and G395M/F290LP grating/filter pairs, providing a  wavelength coverage of 1 -- 5.3 $\mu$m.
Each configuration has a total exposure time of 3063.667 s.
We adopt the NIRSpec data produced by the CEERS collaboration using the STScI JWST Calibration Pipeline\footnote{\url{https://github.com/spacetelescope/jwst}} \citep{Bushouse2022}.
Specifically, we use the JWST pipeline to perform the standard reductions, including the removal of dark current and bias, flat-fielding, background, photometry, wavelength, and slitloss correction for each exposure.
We also perform additional reductions to remove the $1/f$ noise and the snowballs.
The 2D spectra of each target are then rectified and combined to generate the final 2D spectra.
The details of the data reduction are presented in \citet[][]{ArrabalHaro2023} and Arrabal Haro et al. \textit{in prep.} 

The JADES M-Grating data are obtained with the G140M/F070LP, G235M/F170LP, and G395M/F290LP grating/filter pairs, providing a  wavelength coverage of 0.7 -- 5.3 $\mu$m.
The JADES NIRSpec observation is split into three visits and the exposure time for each configuration for each visit is 8,315.667 s.
The single object might be observed in multiple visits and thus, the exposure time for individual objects could be 8,315.667, 16,631.334, or 24,947.0 s.
The JADES NIRSpec data used in this work is downloaded from the MAST HLSP archive\footnote{\url{https://archive.stsci.edu/hlsp/jades}}. 
It is reduced by the JADES collaboration using the NIRSpec GTO collaboration pipeline (Carniani et al. \textit{in prep}).
We refer the readers to \citet{Bunker2023} for details.

We note that the flux units of the CEERS 2D spectral data products are MJy, while the flux units of JADES data products are erg s$^{-1}$ cm$^{-2}$ \AA$^{-1}$.
To unify the flux units of data used in this work, we convert the JADES data to the flux units of MJy.

\subsection{Redshift Determination}

Here we use \texttt{Marz}\footnote{\url{http://samreay.github.io/Marz}} \citep{Hinton2016Marz} to estimate the redshifts of CEERS objects.
\texttt{Marz} employs a line-matching algorithm \citep{Baldry2014}, which cross-correlates both the emission- and absorption-line features of the template and the observed spectrum.
This matches our requirements well as the strong emission lines are usually the only feature of the reionization-era galaxies.

The templates used in \texttt{Marz} include stars, absorption-line galaxies, emission-line galaxies, and quasars. However, since \texttt{Marz} is designed for ground-based optical surveys, the typical rest-frame near-infrared emission lines are not covered in those templates.
To enable the redshift determination for both low- and high-redshift galaxies, we use \texttt{Bagpipes} \citep{Carnall2018} to generate a template with wavelength coverage from 3400 \AA\ to 20000 \AA. 
We adopt a star-formation history including two exponentially-decaying starburst activities with an earlier starburst to form the majority of stellar masses and a late starburst to produce strong emission lines. 
This template shows significant Balmer series, \foiii~$\lambda\lambda4959,5007$ doublets, Paschen-series, \hei~$\lambda10830$, and \siii~$\lambda9530$ lines.
Since the line-matching algorithm is not sensitive to the relative strength between emission lines, altering the \texttt{Bagpipes} input does not significantly change the redshift determination.

To generate the input one-dimensional (1D) spectra, we combine the short- (G140M/F100LP), medium- (G235M/F170LP), and long-wavelength (G395M/F290LP) CEERS 1D spectra by resampling them to a common wavelength grid with the wavelength interval of 6 \AA.
For the wavelength ranges that are covered by two adjacent grating/filter pairs, we adopt the average of two spectra.
In most cases, clear emission lines are detected and an unambiguous redshift can be determined. 
However, if the 1D spectra are too noisy, we visually inspect the 2D spectra to identify the potential emission lines and then use \texttt{Marz} to estimate redshift with the identified emission lines.
If the redshift cannot be estimated, we exclude the object from the final sample.

The redshift estimated by the \texttt{Marz} can sometimes be biased by the noise spikes. 
Thus, we refine the redshifts of high-redshift galaxies by fitting Gaussian profiles to \ha, \hb, and/or \foiii~$\lambda\lambda4959,5007$ lines, whichever is applicable, simultaneously with the \texttt{Marz} redshift as a prior and the wavelengths tied based on the vacuum wavelengths. 
For the JADES objects, we adopt the redshifts measured by the JADES collaboration \citep{Bunker2023}.
Finally, we identified 60 $z>5$ galaxies from the CEERS and 36 $z>5$ galaxies from the JADES with the redshift uncertainties of 30 -- 60 \kms.

\subsection{Sample}

\begin{figure}
    \centering
    \includegraphics[width=3.3in]{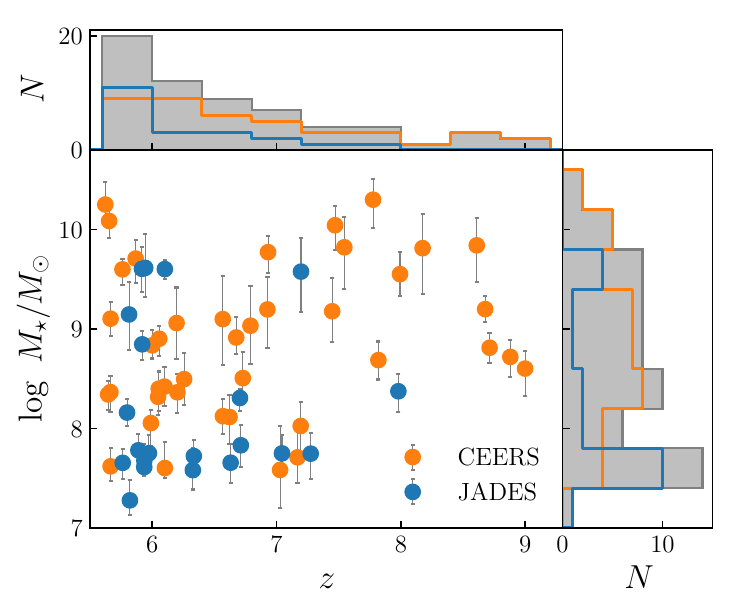}
    \caption{ Redshift versus stellar mass diagram of the galaxies used in this work.
    The orange and blue circles represent the CEERS and JADES samples, respectively.
    We also present the distributions of the redshift and stellar mass of the sample as the histograms in the top and the bottom right panels.
    The gray histograms indicate the distributions of the full sample, while the median redshift is 6.33 and the median stellar mass is $10^{8.55}\ M_\odot$.}
    \label{fig:zdist}
\end{figure}

Because we are interested in the rest-frame UV nebular lines, we select the galaxies of which the \civ\ to \ciii\ are covered by the M-Grating spectra. 
Further, we constrain the galaxies of which the \foiii~$\lambda\lambda4959,5007$ are also covered as these lines are critical to determining the galaxy properties. 
These criteria give a redshift range of $5.6 < z < 9.0$.
Additionally, we visually examine the spectra and remove two galaxies that are severely contaminated by artifacts.
In total, 42 and 21 galaxies are selected from CEERS and JADES, respectively.
We present their properties in Table \ref{tab:4}.

In Figure \ref{fig:zdist}, we present the redshift and the stellar mass of the final sample. The stellar mass is estimated by using \texttt{Bagpipes} \citep{Carnall2018} to fit their multi-wavelength photometry data following the description in \citet{Papovich2023}. 
The redshift is fixed to be the spectroscopic redshift.
We used a star-formation history that follows a delayed-$\tau$ model, where SFR $\sim t\times \exp(-t/\tau)$ with $0.1 < \tau / \mathrm{Gyr} < 10$ and where the age ($t$) is in the range 1 -- 2000~Myr.  The metallicity is allowed to span 0 -- 2.5~$Z_\odot$.  We allow for dust attenuation following \citet{Calzetti2000} with $A(V)$ in the range 0.0 to 5.0~mag.  We include nebular emission with the metallicity of the gas equal to that of the stellar populations, and an ionization parameter,  $\log U$ in the range $-4$ to $-1$. We also allow the nebular escape fraction to span from $10^{-4}$ to 1 (following \citealt{Cole2024}).  We use a linear prior on all parameters except the escape fraction, which uses a log-linear prior.   

The stellar masses of the final sample cover from $10^7$ -- $10^{11}$ $M_\odot$, while on average the JADES galaxies have lower mass compared to the CEERS galaxies.
This is because the CEERS survey is relatively shallower and wider than the JADES survey, and thus preferentially focuses on the brighter galaxies.
The median redshift and median stellar mass of the full sample are 6.33 and $10^{8.55}\ M_\odot$, respectively.

\section{Composite Spectrum and Emission Lines} \label{sec:stack}

\subsection{Constructing Composite Spectrum}

\begin{figure*}
    \centering
    \includegraphics[width=0.95\textwidth]{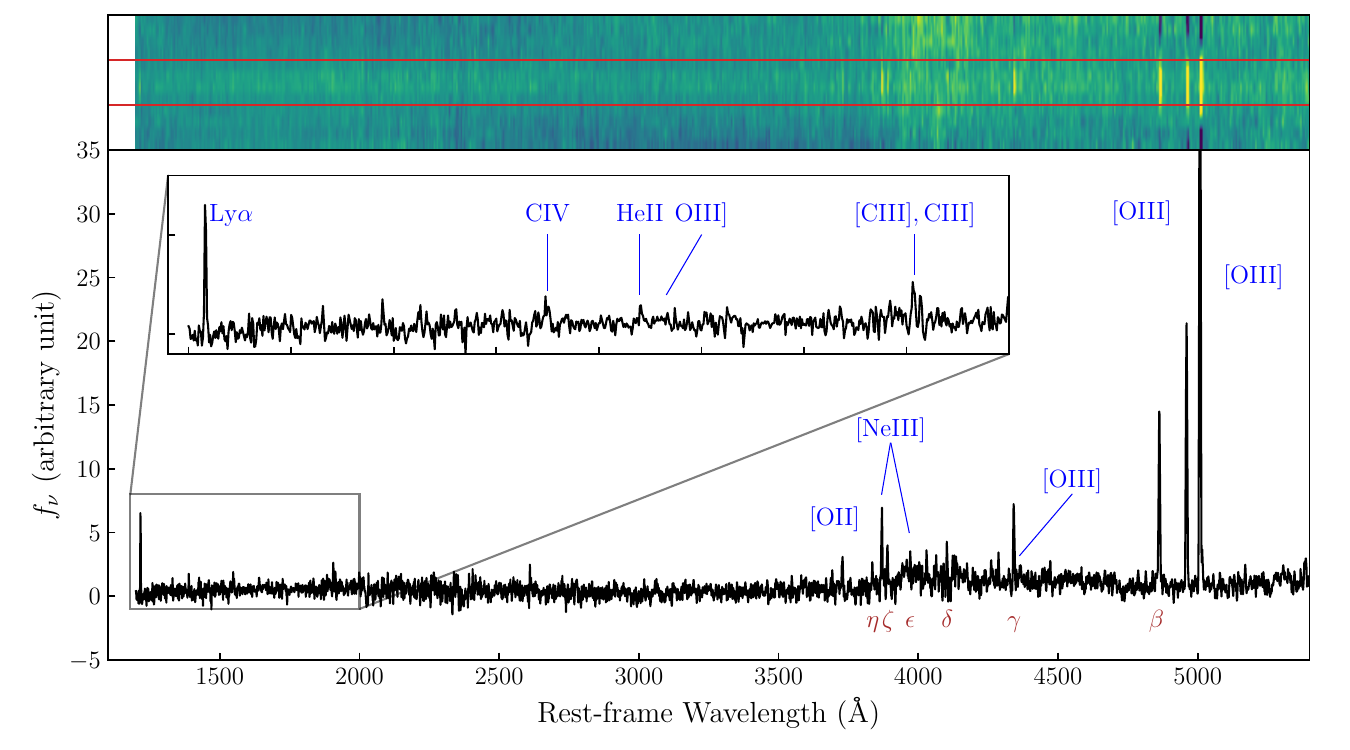}
    \caption{The 2D (top) and 1D (bottom) composite spectra of 63 galaxies at 1200 \AA\ $<\lambda<$ 5400 \AA\ from CEERS and JADES surveys. 
    The 1D composite spectrum is binned by 2 wavelength intervals (i.e., 1 \AA) for better illustration.
    We use blue to label the emission lines detected in this work and use Greek symbols to indicate the hydrogen Balmer lines. The inset panel shows a zoom-in of the wavelength range of 1200 -- 2000 \AA.
    }
    \label{fig:fullspec}
\end{figure*}

\begin{figure*}
    \centering
    \includegraphics[width=0.3\textwidth]{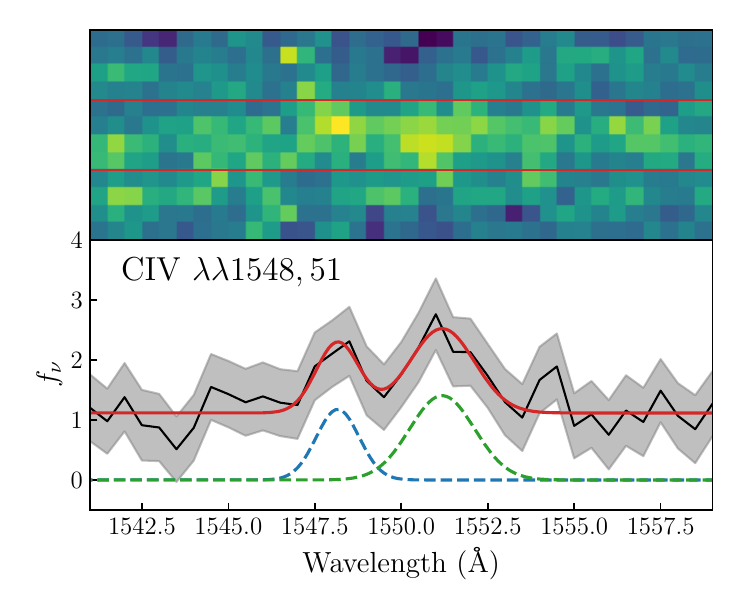}
    \includegraphics[width=0.3\textwidth]{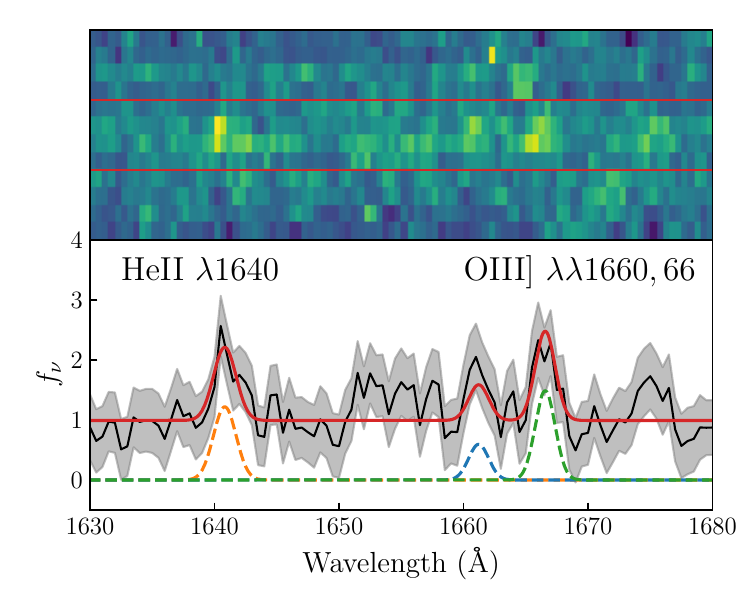}
    \includegraphics[width=0.3\textwidth]{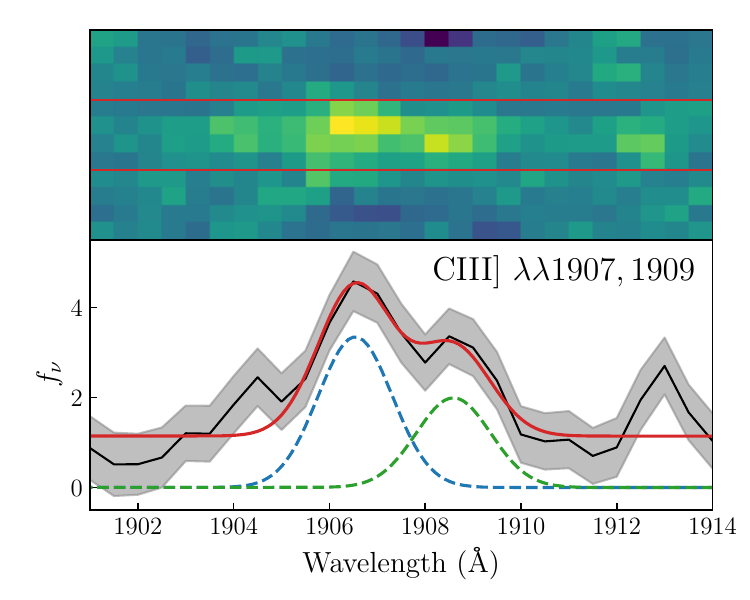}\\
    \includegraphics[width=0.3\textwidth]{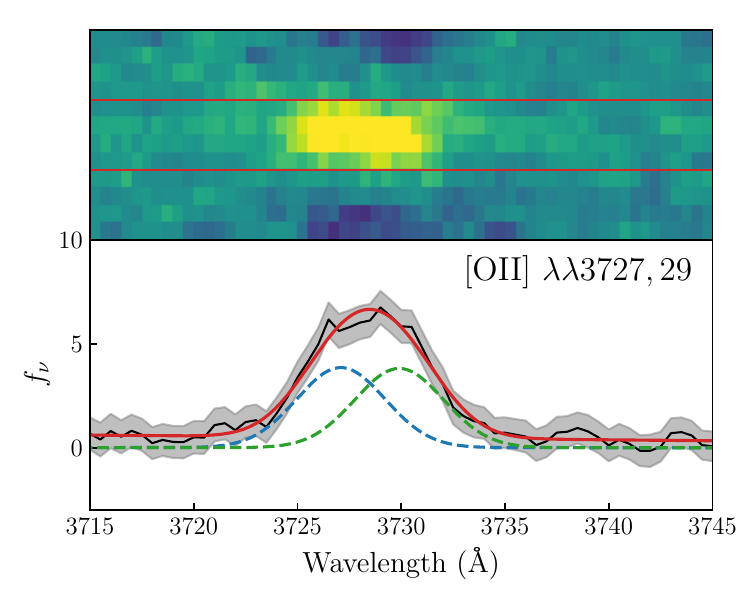}
    \includegraphics[width=0.3\textwidth]{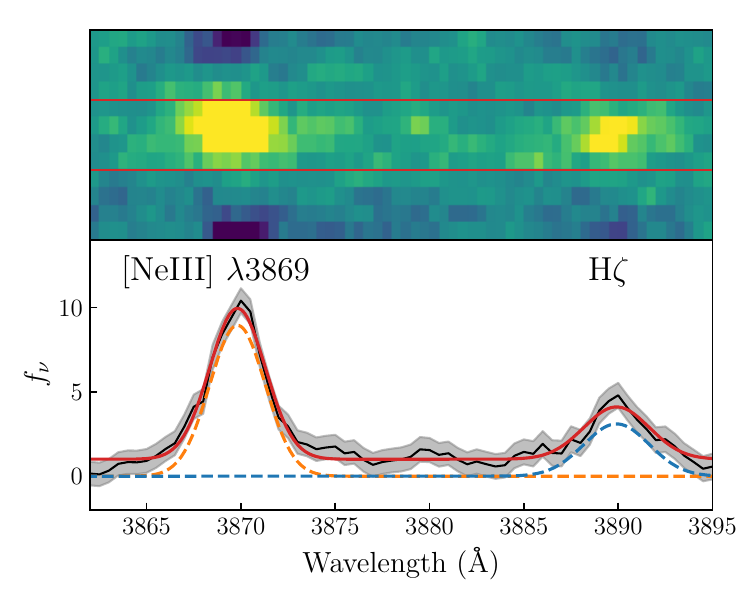}
    \includegraphics[width=0.3\textwidth]{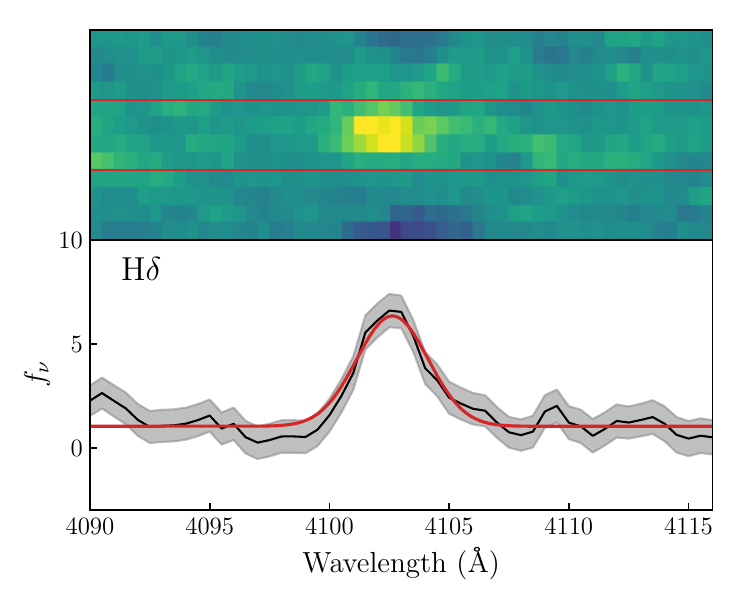}\\
    \includegraphics[width=0.3\textwidth]{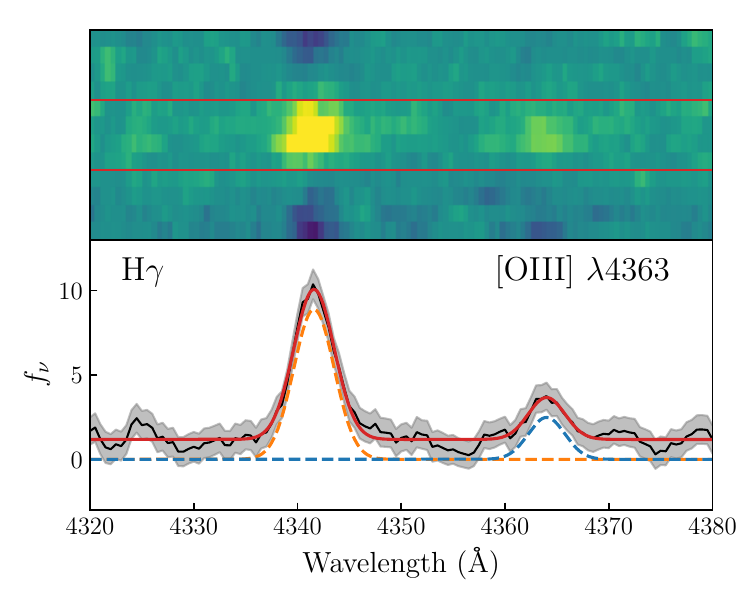}
    \includegraphics[width=0.3\textwidth]{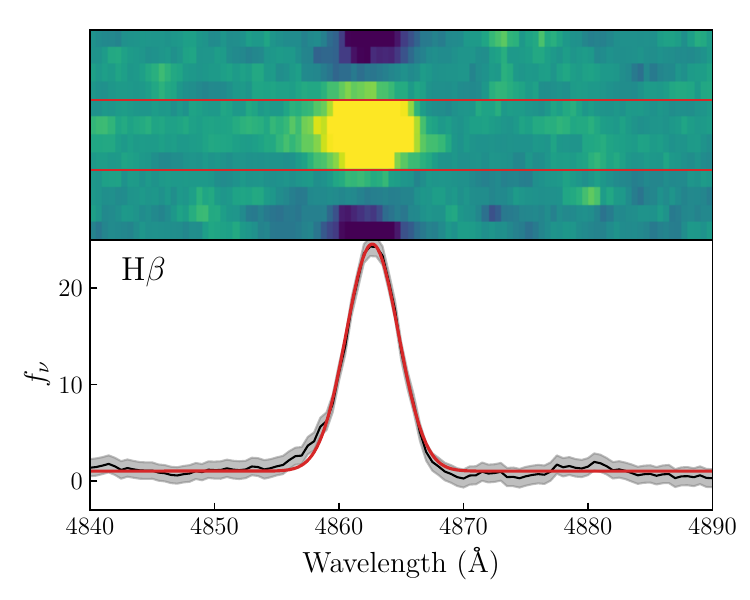}
    \includegraphics[width=0.3\textwidth]{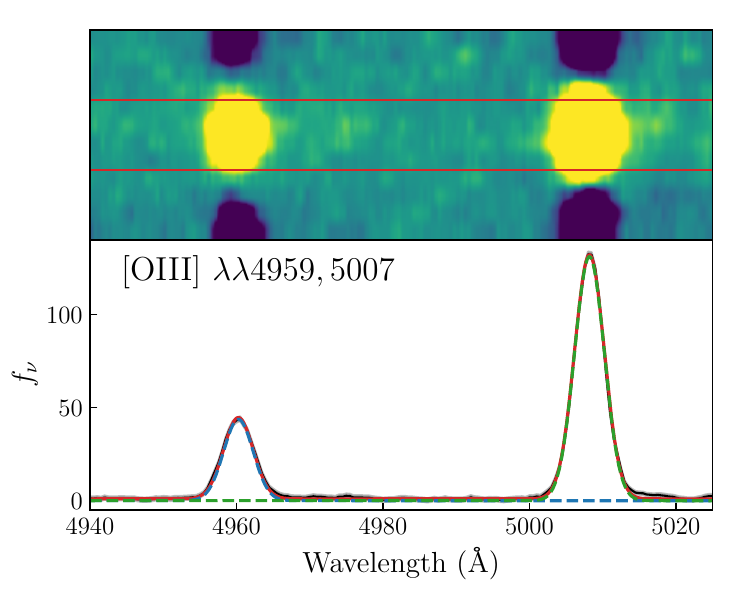}
    \caption{The 2D and unbinned 1D spectra of UV and optical emission lines. In the top panel of each sub-figure, we present the 2D spectra and the extraction window indicated by the two red lines. In the bottom panel of each sub-figure, we present the 1D spectra (black). We use the gray shade to indicate the $1\sigma$ error. The red solid line indicates the best-fit spectra and the orange, blue, or green dashed lines indicate the different components of the best-fit model.}
    \label{fig:emlines}
\end{figure*}

We construct the composite spectrum by median-stacking the normalized 2D spectra of our sample.
Compared with the mean-stacking method, the median-stacking method can avoid the spectral features being dominated by the few outliers in the sample.
In addition, the 2D-stacking method allows us to inspect the composite 2D spectrum to verify the reliability of emission lines.

For each observed spectrum, we first extract a cutout from the 2D spectrum using a 12-pixel window in the spatial direction (corresponding to 1\farcs2) centered on the local maximum of \foiii~$\lambda5007$.
In cases where \foiii\ lacks the spectral coverage, we use \ha, and if both \foiii\ and \ha\ are unavailable, we adopt \hb.
We then use cubic spline interpolation to shift the spectrum to the rest frame on a common wavelength grid, which covers 1200 -- 6000 \AA\ with a wavelength interval of 0.5 \AA. 
An associated variance image is generated based on the error extension and is also shifted to the common wavelength grid.
At the same time, we generate a mask image to mask the wavelength range that falls on the NIRSpec detector gaps. 

We normalize each spectrum based on the galaxy's broadband photometry. 
We adopt the JWST/F150W band to avoid contamination from strong emission lines and \lya\ breaks. 
If JWST/F150W photometry is not available, we adopt the HST/F160W band from CANDELS \citep{Grogin2011,Koekemoer2011}. 
Since our objects span a wide redshift range, the JWST/F150W and HST/F160W bands cover different rest-frame wavelengths.
To mitigate this effect, we adopt the rest-frame 1500 \AA\ magnitudes as the normalization factors, which are converted from their JWST/F150W or HST/F160W photometry with a UV slope of $f_\lambda = \lambda^{-2}$.
Our assumption on the UV slope is consistent with the recent observations of reionization-era galaxies \citep{Cullen2023}. 
In this step, the variance images are also scaled accordingly.
We do not attempt to correct the spectra for the dust attenuation here because we cannot reliably measure the Balmer decrement for all the galaxies in the sample (see Section \ref{sec:dust}).

Although the median-stacking method is less prone to outliers, the outliers still potentially bias the composite spectrum and produce artificial signals. 
By examining the 2D spectra, we find that the outliers usually have very large values compared to other pixels.
Thus, we utilize a $10\sigma$-clipping method to effectively remove the extreme outliers, where the $\sigma$ is the median absolute deviation of all pixels used in the stack, and use the median-stacking method to generate the composite 2D spectrum.
Meanwhile, the error spectrum is constructed by propagating the variance spectrum of each object\footnote{The error of median-stacking is a factor of $\sqrt{\pi/2}\approx1.25$ larger than the error of mean-stacking \citep{Gruen2014}.}.

In the top panel of Figure \ref{fig:fullspec}, we present the full 2D composite spectrum derived in this work. 
The strong emission lines can be easily recognized, along with the faint continuum.
Those optical emission lines are extended in spatial direction and are always accompanied by two dark stripes on both sides, which are caused by the nodded background subtraction.
The weaker UV emission lines are fainter.
Nonetheless, the zoom-in 2D spectra in Figure \ref{fig:emlines} unambiguously reveal the detections of several UV lines.
To compromise between the S/N of UV emission lines, the spatial extent of optical lines, and the dark stripes due to background subtraction, 
we sum the fluxes within a window of 4 pixels (corresponding to 0.4\arcsec) to extract the 1D spectrum. 
We do not perform additional background subtraction as it has been applied during the raw data reduction. 
In the bottom panels of Figure \ref{fig:fullspec}, we present the full 1D composite spectrum and the zoom-in 1D spectrum of the UV and optical emission lines used in this work.

\subsection{Line Measurements} \label{sec:measurements}

The emission line fluxes are measured by fitting the spectra with Gaussian profiles and a constant continuum $f_{\nu,cont}$. 
The optical emission lines are sufficiently bright that varying the continuum level does not significantly change the flux measurements.
Therefore, the continuum level is a free parameter when fitting the optical lines.
However, the UV lines are much weaker and the flux measurements are sensitive to the continuum determination.
To mitigate the uncertainty introduced by the continuum determination, we mask the emission lines and adopt the $3\sigma$-clipped median within a wavelength range of $-30$\AA\ to $+30$ \AA\ of emission lines as the continuum levels for UV lines.

We also adopt different recipes for Gaussian components of the UV and optical emission lines:
\begin{itemize}
    \item \civ~$\lambda\lambda1548,1551$: Since \civ\ is resonance lines and sometimes show P-Cygni profiles in local galaxies \citep{Mingozzi2022}, the widths and line centers are allowed to vary.
    
    \item \fciii\ $\lambda$1907,\ciii\ $\lambda$1909 (hereafter \ciii\ $\lambda\lambda1907,1909$): The line centers of \ciii\ $\lambda\lambda1907,1909$ are fixed to their vacuum wavelengths and the line widths are set to be same.

    \item \heii~$\lambda1640$, \oiii\ $\lambda\lambda1660,1666$: Since these three emission lines are marginally detected, we fix their line centers based on their vacuum wavelengths. We also set the line widths of \oiii\ $\lambda\lambda1660,1666$ to be the same and fix the ratio of \oiii\ $\lambda1666/\lambda1660$ to be the theoretical value of 2.49 \citep{Aggarwal1999}.

    \item \oii~$\lambda\lambda3727,3729$: The \oii\ doublet is resolved in our composite spectrum. To separate the two lines, we fix the line centers of \oii~$\lambda\lambda3727,3729$ lines based on their vacuum wavelengths and set the line widths to be the same.

    \item \foiii~$\lambda\lambda4959,5007$: We set the widths of \foiii\ $\lambda\lambda5007,4959$ to be same, and fix the ratio of \foiii\ $\lambda4959/\lambda5007$ to be the theoretical value of 2.98 \citep{Storey2000}.
    
    \item Balmer lines: Although \hb\ and \hg\ lines are detected at very high S/N, allowing us to robustly measure their fluxes, the S/Ns of other Balmer lines are much lower. Thus, to reliably measure the fluxes of fainter Balmer lines, we utilize the information from \hb\ line.
    We first use a Gaussian profile with free width, strength, and line center to fit the \hb\ line. 
    We then fix the line widths and centers of the Gaussian profile for other Balmer lines based on \hb.
\end{itemize}

To estimate the uncertainty of the flux measurements, we adopt a Monte Carlo approach by perturbing the line profiles according to their error spectrum. 
We generate 1000 realizations for each line and refit them using the above method. 
Then we adopt the standard deviation of the distribution of flux measurements as the uncertainty.
We do not attempt to measure the continuum and the equivalent widths of those lines, because some systematic background issues shown in the 2D composite spectrum (see Figure \ref{fig:fullspec}) may result in an overestimation of the continuum and underestimation of the equivalent widths.
In Table \ref{tab:1} and Table \ref{tab:2}, we list the fluxes of the emission lines of interest. 

We do not consider the effect of absorption when we measure the Balmer line fluxes because our composite spectrum cannot resolve them. 
However, the contribution of Balmer absorption lines, which relies on the average stellar population of our sample, might not be negligible, in particular to the fluxes of high-order Balmer lines.
To estimate their contribution, we adopt the relation between the equivalent widths of the Balmer emission line and absorption line presented in \citet{Peng2023}.
From the composite spectrum, we derive a lower limit of the equivalent width of the \hg\ line to be $>40$ \AA.
Compared to Figure 4 in \citet{Peng2023}, we find that the equivalent width of the \hg\ absorption line should be $\lesssim 3$ \AA.
Assuming the \hb\ and \hd\ absorption lines have the same equivalent width as \hg\ absorption line \citep{Groves2012}, we expect the contribution of absorption lines to \hb, \hg, and \hd\ to be $<2.2\%$, $<7.5\%$, and $<15\%$, respectively.  We consider these as systematic uncertainties on these emission-line fluxes. 

We further acknowledge several potential issues of flux measurement related to the data reduction. 
First, the flux calibration of the JWST pipeline is based on the pre-launch models, which can result in systematic flux calibration uncertainties in the range of 15\% -- 40\%\footnote{\url{https://jwst-docs.stsci.edu/jwst-calibration-pipeline-caveats/jwst-nirspec-mos-pipeline-caveats}}.
Second, the slit losses of the CEERS and JADES data are corrected by assuming the objects are point-like sources and adopting the pre-launch models.
In addition, since the PSF of JWST varies significantly with the wavelength \citep{deGraaff2023}, the extraction window adopted in this work can also result in different flux losses for UV and optical emission lines.
These three effects are wavelength-dependent and can introduce uncertainties to the absolute flux measurements.
Thus, in this work, we focus on the flux ratios of emission lines with small wavelength separations, which are relatively insensitive to these effects.

\begin{deluxetable}{c c c}
    \tablecaption{Line Fluxes Relative to \hb\ \label{tab:1}}
    \tablehead{
    \colhead{Line} & \colhead{Flux$^a$} & \colhead{Dust-corrected Flux$^b$}
    }
    \startdata
    \civ~$\lambda$1548 & $14.57 \pm 7.99$ & $33.00 \pm 18.08$\\
    \civ~$\lambda$1551 & $26.20 \pm 9.93$ & $59.32 \pm 22.50$\\ 
    \heii~$\lambda$1640 & $19.71 \pm 6.87$ & $42.68 \pm 14.87$\\ 
    \oiii~$\lambda$1660 & $8.28 \pm 1.99$ & $17.76 \pm 4.27$\\
    \oiii~$\lambda$1666 & $20.60 \pm 4.97$ & $44.00 \pm 10.63$\\
    \fciii~$\lambda$1907 & $34.14 \pm 4.98$ & $66.16 \pm 9.64$\\
    \ciii~$\lambda$1909 & $20.28 \pm 4.41$ & $39.37 \pm 9.67$\\   
    \oii~$\lambda$3727 & $28.31 \pm 3.70$ & $34.09 \pm 4.45$\\
    \oii~$\lambda$3729 & $27.97 \pm 3.15$ & $33.67 \pm 3.79$\\
    \neiii~$\lambda$3869 & $42.07 \pm 2.34$ & $49.31 \pm 2.74$\\
    H$\zeta$ & $17.20 \pm 2.11$ & $20.08 \pm 2.46$\\ 
    \hd & $22.55 \pm 1.94$ & $25.35 \pm 2.18$\\ 
    \hg & $43.91 \pm 2.20$ & $47.45 \pm 2.37$\\ 
    \foiii~$\lambda$4363 & $11.74 \pm 2.05$ & $12.64 \pm 2.20$ \\
    \hb & $100 \pm 1.81$ & $100 \pm 1.81$\\ 
    % \hb & $209.39 \pm 3.8$ & $412.30 \pm 7.48$\\ 
    \foiii~$\lambda$4959 & $170.70 \pm 0.74$ & $168.78 \pm 0.73$\\
    \foiii~$\lambda$5007 & $508.84 \pm 2.14$ & $498.27 \pm 2.10$\\ 
    \hline
    \enddata
    \tablecomments{Arbitrary line fluxes measured from the composite spectrum and the dust-corrected composite spectrum.
    \begin{itemize}
        \item[$^a$] The dust-uncorrected line fluxes are presented as the values relative to dust-uncorrected \hb\ flux.
        \item[$^b$] The dust-corrected line fluxes are presented as the values relative to dust-corrected \hb\ flux. We adopt a \citet{Calzetti2000} attenuation law with $R_V=4.05$ and an $E(B-V)_\mathrm{gas}=0.16$ to correct the dust attenuation. For details see Section \ref{sec:dust}.
    \end{itemize}}
\end{deluxetable}

\begin{deluxetable}{c c}
    \tablecaption{Line Ratios \label{tab:2}}
    \tablehead{
    \colhead{Line Pair(s)} & \colhead{Flux Ratio}
    }
    \startdata
    \civ~$\lambda\lambda1548,1551$/\heii~$\lambda1640$ & $2.16\pm1.01$ \\
    \oiii~$\lambda1666$/\heii~$\lambda1640$ & $1.03\pm0.44$ \\
    \ciii~$\lambda\lambda1907,1909$/\heii~$\lambda1640$ & $2.48\pm0.92$ \\
    \ciii~$\lambda\lambda1907,1909$/\oiii~$\lambda1666$ & $2.40\pm0.66$ \\
    \civ$\lambda\lambda1548,1551$~/\ciii~$\lambda\lambda1907,1909$ & $0.874 \pm 0.296$ \\
    \fciii~$\lambda$1907/\ciii~$\lambda$1909 & $1.68 \pm 0.41$ \\
    \oii~$\lambda$3729/\oii~$\lambda$3727 & $1.01 \pm 0.18$ \\
    \neiii~$\lambda3869$/\oii~$\lambda3727,3729$ & $0.727\pm0.071$ \\
    \oiii~$\lambda$1666/\foiii~$\lambda$5007 & $0.0883 \pm 0.0213$ \\
    \foiii~$\lambda$4363/\foiii~$\lambda$5007 & $0.0254 \pm 0.0044$ \\
    \foiii~$\lambda\lambda4959,5007$/\oii~$\lambda3727,3729$ & $9.84 \pm 0.79$ \\
    (\foiii~$\lambda\lambda4959,5007$+\oii~$\lambda3727,3729$)/\hb & $7.35 \pm 0.15$ \\
    \hline
    \enddata
    \tablecomments{Line ratios calculated based on the dust-corrected line fluxes in Table \ref{tab:1}.}
\end{deluxetable}

\section{Results} \label{sec:results}
In this section, we use the line ratios measured in Section~\ref{sec:measurements} to derive the average physical properties of the nebular gas in these reionization-era galaxies, including the dust attenuation, the electron density, the electron temperature, and chemical abundance, as shown in Table \ref{tab:3}.

\subsection{Dust attenuation} $\label{sec:dust}$
Correcting for dust attenuation is necessary when inferring galaxy properties from line ratios, especially lines widely separated in wavelength.  
The dust attenuation is commonly estimated by comparing the observed Balmer line ratios to the theoretical ratios.
To reliably measure the Balmer line ratios, it is required to detect at least two Balmer lines at high S/N ratios.
However, most individual galaxy spectra in this work can not provide sufficiently high-S/N Balmer lines for more than a single line ratio.
Thus, we do not correct the dust attenuation for individual galaxies. Instead, we adopt the Balmer decrement measured from the composite spectrum to correct the dust attenuation.
Here we use the ratios of \hb/\hg\ and \hb/\hd, and assume a \citet{Calzetti2000} attenuation law with $R_V=4.05$.

We calculate the intrinsic \hb/\hg\ and \hb/\hd\ ratios to be 2.11 and 3.80 using \texttt{pyneb}\footnote{\url{https://github.com/Morisset/PyNeb_devel}} \citep{Luridiana2015} assuming a Case B recombination\footnote{Comparing with Case A recombination, Case B is a better approximation in the optically-thick ISM as the photons released by free-to-ground recombination can immediately reionize a nearby hydrogen atom \citep[see also][]{Yung2020}.
The intrinsic Balmer line ratios are relatively insensitive to the assumed temperature and density. For example, assuming $T_e$ of 20000 K and $n_e$ of 100 cm$^{-3}$ only decreases the line ratios by $\sim0.5\%$.} with temperature $T_e$ of 17000 K and density $n_e$ of 500 cm$^{-3}$.
The temperature and the density adopted here are consistent with our measurements based on the emission line ratios (see Section~\ref{sec:temp}).
Comparing the observed ratio to the intrinsic ratio, we obtain the dust attenuation of $E(B-V)_\mathrm{gas}=0.16^{+0.10}_{-0.11}$ and $0.21^{+0.12}_{-0.13}$ from \hb/\hg\ and \hb/\hd, which agree within $1\sigma$ uncertainty.
We stress again that the contribution of Balmer absorption lines to the emission-line fluxes is not considered here, because they are not resolved in our composite spectrum (see Section \ref{sec:measurements}).
If we adopt the upper limit of the estimation for Balmer line fluxes estimated in Section \ref{sec:measurements}, we obtain the dust attenuation of $E(B-V)_\mathrm{gas}=0.06^{+0.10}_{-0.06}$ and $0.04^{+0.10}_{-0.04}$ from \hb/\hg\ and \hb/\hd, respectively.
The estimation of dust attenuation from \hb/\hd\ is more affected by the underlying absorption.
We, therefore, adopt $E(B-V)_\mathrm{gas} = 0.16$ from \hb/\hg\ and correct the dust attenuation of emission lines using this value. The uncertainty from dust attenuation is roughly $\pm0.1$~mag, which translates to 10\% at 3727~\AA\ and 18\% at 1500~\AA.
The dust-corrected fluxes and line ratios are presented in Table \ref{tab:1} and \ref{tab:2}.

\subsection{Temperature and Electron Density} \label{sec:temp}

\begin{figure*}
    \centering
    \includegraphics[width=3.2in]{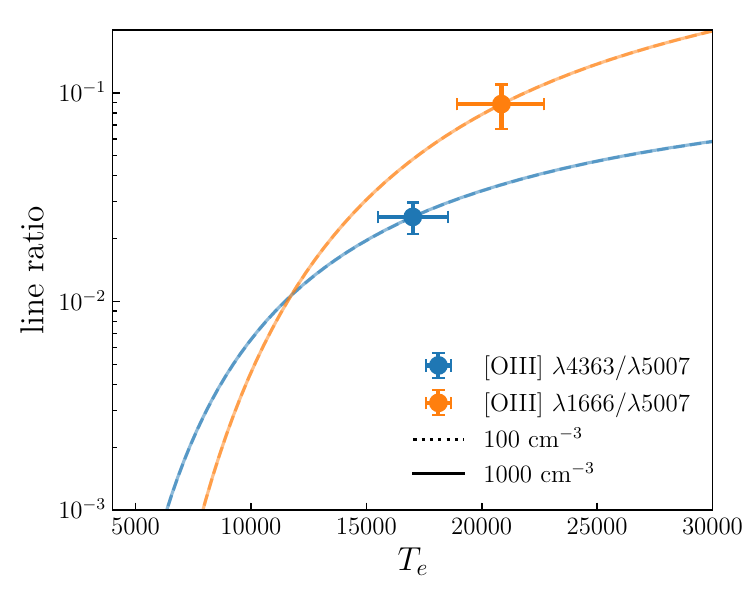}
    \includegraphics[width=3.2in]{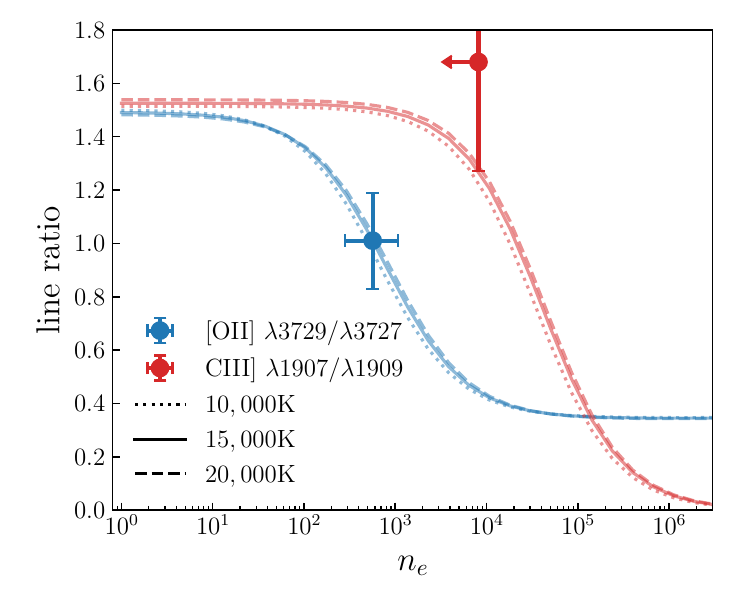}
    \caption{The temperature (left) and electron density (right) diagnostics. 
    We mark the observed line ratios as solid circles.
    The left panel shows the \foiii~$\lambda4363/\lambda5007$ (blue) and \oiii~$\lambda1666/$\foiii~$\lambda5007$ (orange) ratios as a function of temperature.
    We present the line ratios for the electron densities of 100 and 1000 cm$^{-3}$ as the dotted and solid lines. The relations are nearly unchanged from 100 to 1000 cm$^{-3}$.
    The right panel shows the \oii~$\lambda3729/\lambda3727$ (blue) and \ciii~$\lambda1907/\lambda1909$ (red) as a function of electron density.
    We use dotted, solid, and dashed lines to indicate the line ratios for the temperatures of 10000, 15000, and 20000 K, respectively.}
    \label{fig:properties}
\end{figure*}

The electron temperature and density are calculated using \texttt{pyneb} and the line ratios observed in this work.
For most ions in this work, we adopt the default atomic data in \texttt{pyneb} while assuming a five-level atom model.
For \oiii\ $\lambda\lambda1660,1666$, we adopt the data from \citet{Aggarwal1999} which calculates the collision strengths for the necessary six-level atom.
For illustration, we use \texttt{pyneb} to calculate the line ratios as a function of temperature and density and present in Figure \ref{fig:properties}.

We first use \oiii\ $\lambda1666$/\foiii\ $\lambda5007$ and \foiii\ $\lambda4363$/\foiii\ $\lambda5007$ to calculate the temperature as it is insensitive to the electron density.
In the left panel of Figure \ref{fig:properties}, we present the line ratio-temperature relations for different electron densities and the relations are nearly unchanged from 100 to 1000 cm$^{-3}$.
We derive temperatures of $17000^{+1500}_{-1500}$ K and $20800^{+1900}_{-1900}$ K from \foiii~$\lambda4363$/\foiii~$\lambda5007$ and \oiii~$\lambda1666$/\foiii~$\lambda5007$, respectively.
We note a small difference between these two temperatures. 
This might indicate the UV dust attenuation of high-redshift galaxies deviates from our dust correction constrained by optical Balmer lines. For example, if the UV-dust attenuation is underestimated from the Balmer-line ratios, the intrinsic \oiii~$\lambda1666$/\foiii~$\lambda5007$ ratio would increase, bringing it in line with the measurement from \foiii\ $\lambda4363$/\foiii\ $\lambda5007$.  This would require an increase in the \oiiiuv~$\lambda$1666 flux of about 40\%, corresponding to a magnitude correction of $A$(1666~\AA) $\approx 0.3-0.4$~mag. 
On the other hand, as mentioned in Section~ \ref{sec:measurements}, the wavelength-dependent flux losses can result in an overestimation of \oiii~$\lambda1666$/\foiii~$\lambda5007$ and thus, overestimate the temperature.
Thus, in this work, we adopt the temperature measurement from \foiii~$\lambda4363$/\foiii~$\lambda5007$, which is much less sensitive to these effects.  Future studies of JWST spectra of high-redshift galaxies with accurate flux calibration can test the details of the nebular temperature \textit{and} the UV--optical dust attenuation.   

We then use \oii~$\lambda3729/\lambda3727$ and \ciii\ $\lambda1907/\lambda1909$ lines ratios to calculate the electron densities.
As shown in the right panel of Figure~\ref{fig:properties}, the two line ratios probe different electron density ranges:
the \oii~$\lambda3729/\lambda3727$ is sensitive to 10 -- $10^4$ cm$^{-3}$ while the \ciii$\lambda1907/\lambda1909$ is sensitive $10^3$ -- $10^6$ cm$^{-3}$.
These line ratios show little dependence on the temperature, which is relatively small compared to the uncertainties of observed line ratios. 
Thus, in this work, we assume the \foiii\ temperature and do not consider the uncertainties introduced by the temperature uncertainty.
We obtain electron densities of $570^{+510}_{-290}$ and $<8600$ cm$^{-3}$ from \oii\ and \ciii, respectively.
The electron density derived from \oii\ is consistent with the measurements for individual galaxies at similar redshift \citep{Isobe2023}.
Because the electron density measured from \ciii\ only provides an upper limit, we adopt the measurement from \oii\ in this work.

\begin{deluxetable*}{l c c r}
    \tablecaption{Average Galaxy Properties \label{tab:3}}
    \tablehead{
    \colhead{Property} & \colhead{Line Ratio Used} & \colhead{Value} & \colhead{Note}
    }
    \startdata
    $z_\mathrm{med}$ & & 6.33 &  \\
    \hline
    $E(B-V)_\mathrm{gas}$ & \hb/\hg & $0.16^{+0.10}_{-0.11}$ & Section \ref{sec:dust}\\
    & \hb/\hd & $0.21^{+0.12}_{-0.13}$ & Section \ref{sec:dust}\\
    \hline
    $T_e(\mathrm{[O\ \textsc{iii}]})$ & \foiii\ $\lambda4363$/\foiii\ $\lambda5007$ & $17000^{+1500}_{-1500}$ & Section \ref{sec:temp}\\
    & \oiii\ $\lambda1666$/\foiii\ $\lambda5007$ & $20800^{+1900}_{-1900}$ &  Section \ref{sec:temp}\\
    \hline
    $n_e(\mathrm{[O\ \textsc{ii}]})$ & \oii\ $
    \lambda3729/\lambda 3727$ & $570^{+510}_{-290}$ & Section \ref{sec:temp}\\
    $n_e(\mathrm{[C\ \textsc{iii}]})$& \ciii\ $\lambda1907/\lambda1909$ & $<8600$ & Section \ref{sec:temp} \\
    \hline
    % $\log(P/k)$ & $6.99^{+0.28}_{-0.32}$ & \\
    $\log(U)$ & \foiii\ $\lambda\lambda4959,5007$/\oii\ $\lambda\lambda3727,3729$ & $-2.18^{+0.03}_{-0.03}$ & Section \ref{sec:coa} \\
    \hline
    $12+\log \mathrm{(O/H)}$ & \foiii\ $\lambda\lambda4959,5007$/\hb, \oii\ $\lambda\lambda3727,3729$/\hb & $7.672\pm0.083$ & Section \ref{sec:oa} \\
    \hline
    $\log \mathrm{(C/O)}$ & \ciii\ $\lambda1909$/\oiii\ $\lambda1666$, $\log(U)$ & $-0.87^{+0.13}_{-0.10}$ & Section \ref{sec:coa} \\
    & \ciii\ $\lambda1909$/\oiii\ $\lambda1666$, \civ\ $\lambda\lambda1548,1551$/\ciii\ $\lambda\lambda1907,1909$ & $-0.70^{+0.13}_{-0.10}$ & Section \ref{sec:coa} \\
    \hline
    \enddata
    \tablecomments{ Average galaxy properties estimated from the line ratios derived in this work.}
\end{deluxetable*}

\subsection{Chemical Abundances}

\subsubsection{Oxygen abundance} \label{sec:oa}

With the temperature and electron density measured above, we are able to use the direct $T_e$ method to determine the oxygen abundance, 12 + log(O/H).
We adopt the parameterizations for O$^{+}$ and O$^{2+}$ from \citet{Peng2023}, which are optimized for the temperature range of 7000 -- 25000 K and the density range of 10 -- 1000 cm$^{-3}$.
Because the temperature-sensitive \oii\ lines (e.g., \oii\ $\lambda\lambda7320,7330$) are not covered in our composite spectrum, we estimate the \oii\ temperature $T_e$\oii\ from the \foiii\ temperature $T_e$\foiii\ using the relation in \citet{Arellano-Cordova2020}.
We obtain 12 + log(O/H) $=7.672\pm0.083$ from the composite spectrum, corresponding to 10\% of the Solar value \citep[8.69;][]{Asplund2021}.

\subsubsection{Carbon-to-Oxygen abundance ratio} \label{sec:coa}
We determine the C/O abundance ratio using the C$^{2+}$/O$^{2+}$ ratio and apply a carbon ionization correction factor (ICF) to account for the contribution of C$^{3+}$ ions:
\begin{equation}\label{eqn:c_o}
    \frac{\mathrm{C}}{\mathrm{O}} = \frac{\mathrm{C^{2+}}}{\mathrm{O^{2+}}} \times \left[ \frac{X(\mathrm{C^{2+}})}{X(\mathrm{O^{2+}})} \right]^{-1} = \frac{\mathrm{C^{2+}}}{\mathrm{O^{2+}}} \times \mathrm{ICF} ,
\end{equation}
where $X(\mathrm{C^{2+}})$ and $X(\mathrm{O^{2+}})$ are the $\mathrm{C^{2+}}$ and $\mathrm{O^{2+}}$ column fraction, respectively.
The C$^{2+}$/O$^{2+}$ abundance is calculated using the \ciii~$\lambda1909$/\oiii~$\lambda1666$ ratio and the corresponding emissivity ratio calculated by \texttt{pyneb}. 
We adopt the temperature $T_e=17000^{+1500}_{-1500}$ K and electron density $n_e=570^{+510}_{-290}$ cm$^{-3}$ obtained in Section \ref{sec:temp} and find log(C$^{2+}$/O$^{2+}$) to be $-0.88^{+0.13}_{-0.10}$, significantly lower than the Solar value, $\log(\mathrm{C/O}) = -0.23$ \citep{Asplund2021}.
We note that here we assume the C$^{2+}$ zone has the same temperature and electron density as the O$^{2+}$ zone.
If we adopt the upper limit of electron density (8600 cm$^{-3}$) measured from \ciii~$\lambda1907/\lambda1909$ line flux ratio for the C$^{2+}$ zone, the log(C$^{2+}$/O$^{2+}$) changes by only $\sim0.005$ dex.

The contribution of C$^{3+}$ ions might not be  negligible, as the \civ\ emission lines are visible in the composite spectrum. 
We use the photoionization model-derived ICF from \citet{Berg2019} to correct the C$^{3+}$ abundance. 
\citet{Berg2019} use \texttt{CLOUDY} \citep{Ferland2013} and BPASS \citep{Eldridge2016,Stanway2016} to estimate the ICF as a function of ionization parameter. 
We use the relations in \citet{Berg2019} for $Z=0.1\ Z_\odot$ to convert the \foiii/\oii\ (O32) ratio to the ionization parameter and the ICF.
We find that the ionization parameter is $\log\ U = -2.18^{+0.03}_{-0.03}$ and the ICF for Equation~\ref{eqn:c_o} is $1.026^{+0.008}_{-0.009}$. 
The errors are propagated from the uncertainty of the \foiii/\oii\ ratio and we do not consider the uncertainty from different models, which could be much larger.
Applying this modest ICF to C$^{2+}$/O$^{2+}$, we obtain the C/O abundance ratio of log (C/O) $=-0.87^{+0.13}_{-0.10}$.

On the other hand, the detection of \civ\ allows us to directly estimate the C$^{3+}$/C$^{2+}$ abundance using the \civ/\ciii\ (C43) ratio.
Assuming all the observed \civ\ is from nebular emission and the temperature and density in C$^{3+}$ zone are the same as those in C$^{2+}$, we obtain C$^{3+}$/C$^{2+}$ to be $0.45 \pm 0.13$, similar to the C$^{3+}$/C$^{2+}$ recently estimated for one galaxy at $z=6.23$ \citep{Jones2023}.
This ratio indicates a correction of $1.45 \pm 0.13$ to C$^{2+}$/O$^{2+}$ and leads to a C/O abundance ratio of log (C/O) $=-0.70^{+0.13}_{-0.10}$. 

Interestingly, the C43-based ICF is significantly larger than the O32-based ICF, resulting in different estimations of the C/O abundance ratio.
This discrepancy is also seen in some low-redshift studies \citep{Berg2016,Berg2019}.
As \civ\ can also originate from stellar photospheres, this discrepancy might reflect that a large fraction of \civ\ emission is of stellar origin.
For example, by comparing the stellar population model \texttt{STARBURST99} \citep{Leitherer1999} to the galaxy spectra at $z\sim3.1$ -- 4.6, \citet{Saxena2022} found that on average $\sim25\%$ of the \civ\ flux could be attributed to the stellar origin.
Due to insufficient resolution and relatively low S/N in the rest-frame UV, we are unable to directly determine the stellar contribution of \civ.
If we assume the excess of C43-based ICF is contributed by the stellar \civ\ emission, we can find that 91 -- 95$\%$ of \civ\ flux originates from hot stars, rather than the ionized ISM.

In this work, we adopt the C/O abundance ratio derived using the O32-based ICF to avoid the complexity of \civ\ origins.

\section{Discussion} \label{sec:discussion}

\subsection{Interpretation of the C/O Abundance Ratio} \label{sec:co}

\begin{figure*}
    \centering
    \includegraphics[width=5in]{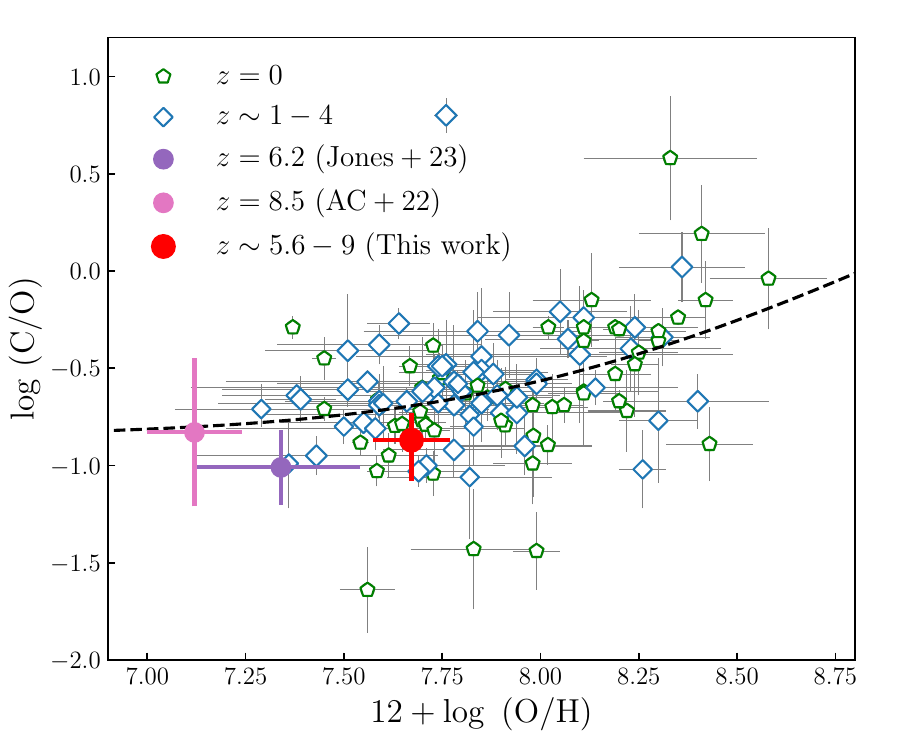}
    \caption{C/O-O/H relationship. We mark our measurement of the composite spectrum as the red circle.
    For comparison, we compile several objects at $z\sim0$, $z\sim1-4$, and $z>6$ from the literature.
    The first two JWST C/O measurements of reionization-era galaxies are presented as pink \citep[][AC+22 in the figure]{Arellano-Cordova2022} and purple \citep{Jones2023} circles.
    The green open pentagons mark the measurements of dwarf galaxies at $z\sim0$ \citep{Berg2016,Berg2019,Pena-Guerrero2017,Senchyna2017,Senchyna2021,Rogers2023}.
    The blue open diamonds mark the measurements of galaxies at $z\sim1-4$ \citep{Erb2010,Christensen2012,James2014,Bayliss2014,Stark2014,Steidel2016,Amorin2017,Berg2018,Mainali2020,Matthee2021,Citro2023,Iani2023,Llerena2023}.
    Motivated by Equation 3 in \citet{Nicholls2017}, we use a similar equation to fit the C/O-O/H relation of $z\sim0$ and $z\sim1-4$ galaxies, and the best-fit curve is indicated by the black dashed line.
    }
    \label{fig:coabundance}
\end{figure*}

The chemical enrichment of galaxies is a powerful probe for studying galaxy evolution. 
The relative abundance of carbon to oxygen (C/O) is particularly of interest because those elements are produced by different nucleosynthetic processes on different timescales \citep[for a review see][]{Maiolino2019}.
Oxygen is primarily produced by short-lived massive stars (with lifetimes of $\sim10$ Myr) and then ejected to the ISM by core-collapse supernovae. 
The strong outflows driven by short-lived massive stars can also preferentially remove oxygen on short timescales prior to C production.
In contrast, carbon enrichment is eventually dominated by the intermediate-mass stars ($2<M/M_\odot<8$) during their Asymptotic Giant Branch (AGB) phase (with lifetimes of $\sim 200$ Myr -- 1 Gyr). 
Consequently, the C/O abundance ratio can indicate the key ingredients of galaxy evolution, such as outflows \citep[e.g.,][]{Yin2011} and star formation history \citep[e.g.,][]{Berg2019,Vincenzo2018}.

In Figure~\ref{fig:coabundance}, we compare our C/O-O/H measurements from the composite spectrum of reionization-era galaxies with other measurements from the literature, including
the first two JWST C/O measurements of galaxies at $z>6$ \citep{Arellano-Cordova2022,Jones2023},
measurements of local dwarf galaxies \citep{Berg2016,Berg2019,Pena-Guerrero2017,Senchyna2017,Senchyna2021,Rogers2023} and intermediate-redshift galaxies \citep[$z\sim1-4$;][]{Erb2010,Christensen2012,James2014,Bayliss2014,Stark2014,Steidel2016,Amorin2017,Berg2018,Mainali2020,Matthee2021,Citro2023,Iani2023,Llerena2022,Llerena2023}.
Although our measurement from the composite spectrum falls within the distribution of $z\sim0$ and $z\sim1-4$ galaxies on the C/O-O/H plane, the C/O and O/H ratios of the composite spectrum tend to be lower than the average values at lower redshifts, implying that the reionization-era galaxies on average are more carbon-poor at a given oxygen abundance.

Quantitatively, we follow \citet{Nicholls2017} to fit the distribution of $z\sim0$ and $z\sim1-4$ galaxies with a simple expression that combines the two origins of carbon enrichment:
\begin{equation}
    \mathrm{log\ (C/O)} = \log\ (10^a+10^{[\mathrm{log\ (O/H)+b}]}).
\end{equation}
The best-fit curve, denoting the average C/O-O/H relation of low-redshift galaxies, is plotted as the black dashed line in Figure \ref{fig:coabundance}.
We obtain $a=-0.85$ and $b=-8.88$.
Compared to the best-fit curve, the C/O ratio of our composite spectrum is lower by $\sim 1\sigma$.

A potentially smaller C/O ratio with moderate oxygen abundance might suggest that reionization-era galaxies are undergoing a rapid build-up of stellar mass, during which the oxygen has been largely released by supernovae but the secondary carbon enrichment from AGB stars has only mildly proceeded. 
Considering that secondary carbon enrichment occurs at $>200$ Myr \citep{Maiolino2019}, this indicates that the majority of the stellar mass is assembled rapidly in these galaxies, on timescales of 100 -- 200  Myr ago.  This is consistent with the short star-forming episodes ($\lesssim 100$~Myr) inferred from the photometric data of galaxies at these redshifts \citep[e.g.,][Papovich, in prep.]{Whitler2023,Caputi2024,Cole2024,Endsley2024}
This is also in alignment with the chemical abundance calculation of the  $z=6.2$ galaxy reported by \citet{Jones2023}, for which the C/O abundances suggest a very young age ($<150$ Myr), consistent with its star formation history inferred from modeling its broad-band photometry.

Another possibility is that reionization-era galaxies have a relatively lower preferential removal of oxygen by outflows\footnote{Because the outflows driven by supernovae or stellar radiation occur before the secondary carbon enrichment from AGB stars, the oxygen is preferentially removed.} compared with low-redshift galaxies with similar oxygen abundances. 
\citet{Berg2019} used a chemical evolution model to demonstrate that the vertical location of the chemical evolutionary track in the C/O-O/H plane is determined by the fraction of oxygen ($X_\mathrm{out}$(O)) removed by the outflows. 
However, this possible origin is at odds with the recent observation of the mass-metallicity relation of reionization-era galaxies.
\citet{Curti2023} found that the high-redshift galaxies deviate from the local fundamental metallicity relation,
which might hint at the oxygen-enhanced outflows.
Consequently, the lower preferential removal of oxygen by outflows is unlikely the case for reionization-era galaxies.

We note again that our measurement significance of a lower C/O ratio in reionization-era galaxies is small and if we adopt the C43-based ICF (i.e., assuming the observed \civ\ is nebular origin), the C/O ratio of our composite spectrum would agree with the low-redshift trend.
Thus, to firmly establish the redshift evolution of C/O ratios,
it is necessary to assemble a larger sample of galaxies with measurements of \civ, \ciii, and \oiii\ at high redshift and discriminate the stellar and nebular contributions to the observed \civ\ lines.

\subsection{Interpretation of the Ionization Properties}

\begin{figure*}
    \centering
    \includegraphics[width=5in]{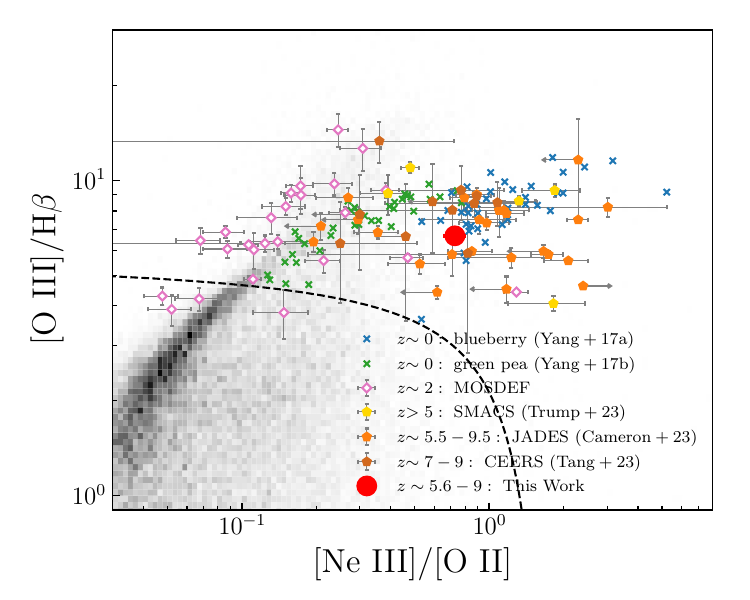}
    \caption{\foiii\ $\lambda\lambda4959,5007$/\hb\ versus \neiii\ $\lambda3869$/\oii\ $\lambda\lambda3727,3729$ diagram. The line ratios measured from this work are marked as the red circle.
    We also plot a series of comparison samples from the literature. 
    The background shows the 2D distribution of $z\sim0$ galaxies from SDSS \citep{Aihara2011} and the black dashed line indicates the SF-AGN separator suggested in \citet{Backhaus2022}.
    The local blueberries \citep{Yang2017b} and green peas \citep{Yang2017} are shown as blue and green crosses. 
    The $z\sim2$ galaxies from the MOSDEF survey \citep{Kriek2015} are shown as pink diamonds.
    The recent JWST measurements of galaxies at $z>5$ from SMACS \citep{Trump2023}, JADES \citep{Cameron2023}, and CEERS \citep{Tang2023} surveys are shown as the yellow, orange, and dark orange pentagons. 
    }
    \label{fig:ne3o2}
\end{figure*}

\begin{figure*}
    \centering
    \includegraphics[width=5in]{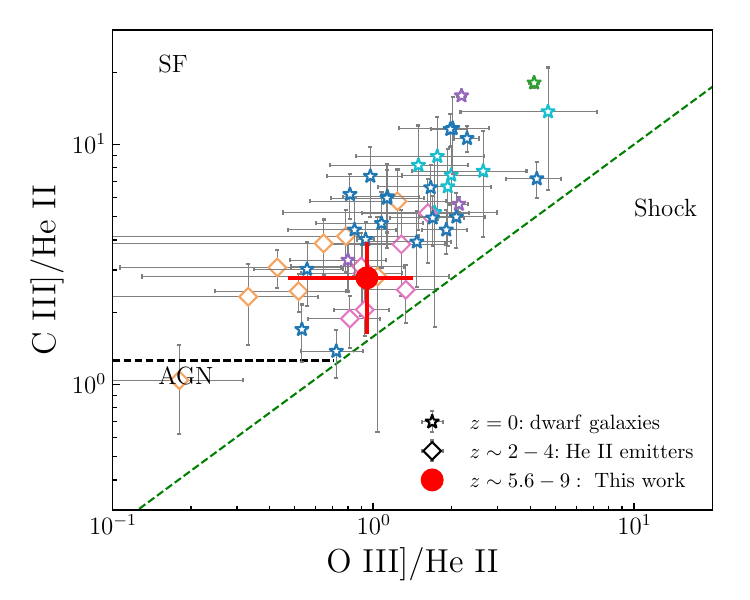}
    \caption{\ciii~$\lambda\lambda1907,1909$/\heii~$\lambda1640$ versus \oiii\ $\lambda1666$/\heii\ $\lambda1640$ diagram. The line ratios measured from this work are marked as the red circle. 
    A series of local galaxies compiled from the literature are marked as purple \citep{Berg2016}, cyan \citep{Berg2019}, green \citep{Senchyna2017}, and blue \citep{Berg2022,Mingozzi2023,Olivier2022} stars. 
    We also include the $z\sim2-4$ \heii\ emitters as pink \citep{Nanayakkara2019} and orange \citep[][]{Saxena2020} diamonds and a $z=6.2$ galaxy as a purple circle \citep[][]{Jones2023}.
    The lines in this figure that separate star-formation (SF)--AGN and SF--shock regions from \citet{Mingozzi2023} are plotted as the black and green dashed lines, respectively.
    Compared to the region spanned by the dwarf stars, our measurement is offset to the lower-left region, close to the two well-known dwarf galaxies with the lowest \oiii/\heii\ ratios, SBS 0335-052 and I Zw 18.
    }
    \label{fig:c3he2}
\end{figure*}

Recent JWST observations have detected the high ionization lines in several individual galaxies, revealing a diverse nature of ionizing sources in high-redshift galaxies from AGNs, possible Population III stars, to young massive stars \citep[e.g.,][]{Fujimoto2023, Larson2023,Cleri2023,Wang2022,Bunker2023b}.
% However, the universal ionizing sources that are common in all galaxies remain puzzles to us. 
However, it remains unclear which ionizing source is the most common in all high-redshift galaxies.
Directly observing the hard ionizing photons from the ionizing source is infeasible. Thus, the flux ratios between emission lines with distinct ionization potentials are frequently used to depict the hardness of the ionizing spectrum. 
In particular, pairs of lines with small wavelength separations are chosen to mitigate the uncertainties from dust attenuation and observation effects.
In this section, we utilize the rest-frame optical and UV line ratio diagnostics to explore the average ionization status in the composite spectrum.

\subsubsection{Optical line diagnostics}

Figure~\ref{fig:ne3o2} presents the \foiii\ $\lambda\lambda4959,5007$/\hb\ versus \neiii\ $\lambda3869$/\oii\ $\lambda\lambda3727,3729$ diagram \citep[O3Hb-Ne3O2;][]{Trouille2011,Zeimann2015} with our measurements shown by the red circle.
We also plot a series of comparison samples, including  $z\sim0$ galaxies from the SDSS MPA-JHU catalog \citep{Aihara2011}, $z\sim0$ dwarf galaxies \citep[including blueberries and green peas from][]{Yang2017b,Yang2017}, $z\sim2$ star-forming galaxies from the MOSDEF survey \citep{Kriek2015}, and reionization-era galaxies from the JWST SMACS, JADES, and CEERS surveys \citep{Trump2023,Cameron2023,Tang2023}.
Our measurements generally agree with the distributions of individual galaxies in previous JWST studies. 

Compared to $z\sim0$ SDSS galaxies and $z\sim2$ MOSDEF galaxies, the \neiii/\oii\ ratio of our composite spectrum is much higher. 
This is consistent with previous studies of $z\gtrsim5.5$ galaxies \citep{Cameron2023,Tang2023,Trump2023}, indicating that the ionizing sources in the reionization-era galaxies can produce a much harder ionizing spectrum than is typical for $z\sim0$ -- 3 galaxies. 
However, we are unable to constrain the type of ionizing source from this diagnostic diagram (dashed line in Figure~ \ref{fig:ne3o2}), because modes of AGN and star formation can recover the region spanned by the composite spectrum and individual $z>5$ galaxies \citep[see Figure~4 in][]{Cleri2023}.
Nonetheless, the local dwarf galaxies (including green peas and blueberries) overlap better with our composite spectrum and the individual galaxies from previous studies. In particular, the blueberries on average have even larger \foiii/\hb\ and \neiii/\oii\ ratios than our composite spectrum.   Therefore, the origin of the ionizing source of \neiii/\oii\ remains uncertain.
Studying the local dwarf galaxies could provide clues to understanding the ionizing source in high-redshift galaxies.

\subsubsection{UV line diagnostics}

Several diagnostic diagrams using UV lines have been proposed in recent works \citep{Feltre2016,Jaskot2016,Nakajima2018,Byler2020,Hirschmann2022} to distinguish between star-formation, AGN, and shock-driven photoionization.
In this work, we adopt the \ciii~$\lambda\lambda1907,1909$/\heii~$\lambda1640$ versus \oiii\ $\lambda1666$/\heii\ $\lambda1640$ diagram (hereafter C3He2-O3He2), which is suggested to show the best classification power among those diagnostics \citep{Mingozzi2023}.
In Figure \ref{fig:c3he2}, we present our measurements along with those from local dwarf galaxies
\citep{Berg2016,Berg2019,Berg2022,Senchyna2017} and \heii\ emitters at $z\sim2-4$ \citep{Nanayakkara2019,Saxena2020}.

We overplot the separators between star formation, AGN, and shocks derived by comparing the ionization models \citep{Gutkin2016,Feltre2016,Alarie2019} with the well-studied local dwarf galaxies \citep{Mingozzi2023}.
The position of the composite spectrum in the C3He2-O3He2 diagram indicates that high-redshift galaxies are dominated by ionization from star formation with a modest significance of $\sim1\sigma$.
\citet{Hirschmann2019} also proposed another set of line separators to distinguish between AGN, star formation, and AGN -- star-formation composites, obtained by coupling \citet{Gutkin2016} and \citet{Feltre2016} models.
However, if we adopt the line separators from \citet[][see their Figure 6]{Hirschmann2019}, our composite spectrum falls in the AGN--star-formation--composite region, suggesting that a large fraction of our sample could host weak AGNs.
However, the local dwarf galaxies and \heii\ emitters are also classified as AGN--star-formation composites. 
Thus, the composite region introduced in \citet{Hirschmann2019} may be more applicable to massive galaxies, where lower-mass galaxies (like local dwarfs and high-redshift galaxies) are more likely to show extreme ionization from star formation instead of having a contribution of weak AGNs. 

Compared to the values of C3He2-O3He2 spanned by the local dwarf galaxies, our high-redshift composite spectrum is offset to the lower-left region, close to the two well-known dwarf galaxies with the lowest \oiii/\heii\ ratios, SBS 0335-052 and I Zw 18.
Our high-redshift composite spectrum overlaps better with the region spanned by the $z\sim2$ -- 4 \heii\ emitters.
The offset from the region spanned by most dwarf galaxies might suggest that reionization-era galaxies generally have harder ionizing radiation and are capable of producing more \heii\ ionizing photons ($>54.4$ eV) than those typical dwarf galaxies.

The two extreme dwarf galaxies, SBS 0335-052 and I Zw 18, might be more similar to the high-redshift galaxies and could provide some hints at the \heii\ ionizing sources.
In particular, the extended \heii\ $\lambda$4686 emission has already been observed in these two galaxies \citep{Kehrig2015,Kehrig2018,RickardsVaught2021}.
The extended \heii\ $\lambda$4686 emission disfavors the hypothesis that an AGN is the dominant \heii\ ionizing source.  X-ray observations would improve the interpretation and rule out other scenarios.  For example, ionization from high-mass X-ray binaries is also disfavored by comparing the observed X-ray luminosity to \heii\ luminosity in SBS 0335-052 \citep{Kehrig2018} and I Zw 18 \citep{Kehrig2021}.
Stellar-population models that include the effects of binary stars and their evolution \citep[BPASS;][]{Eldridge2017,Stanway2018} might be able to explain the observed \heii\ fluxes in the SBS 0335-052  \citep{Kehrig2018}, but these models require extremely low metallicities ($Z=0.0005\ Z_\odot$).
A similar problem is also seen in $z\sim2$ -- 4 \heii\ emitters \citep{Nanayakkara2019,Saxena2020} where the \heii\ equivalent widths are usually underpredicted by the BPASS models \citep{Xiao2018} with more realistic metallicities ($Z=0.005, 0.1, 1\ Z_\odot$). 
Consequently, additional ionizing sources, such as Wolf-Rayet stars and stripped helium stars \citep{Drout2023}, might still be needed to reproduce the observed line strengths.

On the other hand, the offset could also be (partly) attributed to the lower C/O abundance ratio of reionization-era galaxies (see Section \ref{sec:co}), which suppresses the carbon emission \citep{Jaskot2016} and decreases the \ciii/\heii\ ratio.
The overall origin could be a combination of both effects, which is hard to determine yet with our composite spectrum.
Distinguishing between these effects would require high-significance detection of UV lines in a large sample of individual galaxies to establish the evolution track of carbon and oxygen enrichment and determine the statistical offset in the UV diagnostic diagrams.

\subsection{Ionizing photon leakage}

One of the main topics of interest regarding the reionization-era galaxies is to understand how the ionizing photons escape from those galaxies and reionize the Universe.
Due to the heavy attenuation of neutral IGM, it is infeasible to directly measure the Lyman continuum (LyC) escape fraction ($f^\mathrm{esc}_\mathrm{LyC}$) for those galaxies.
A variety of indirect probes have been proposed, including the \lya\ profile, \foiii/\oii\ (O32) ratio, UV absorption lines, \mgii\ emission line, and \civ\ emission line \citep[e.g.,][]{Izotov2016,Chisholm2020,Gazagnes2020,Hu2023,Xu2023,Saxena2022,Schaerer2022}.
We refer the readers to \citet{Flury2022} for a comprehensive discussion of those indirect probes.

The O32 ratio is one of the most frequently used probes to select ionizing-photon leakers as a high O32 ratio might suggest the existence of density-bounded channels through which the ionizing photos can escape. 
Although the correlation between the $f^\mathrm{esc}_\mathrm{LyC}$ and O32 ratio is still under debate \citep{Izotov2018,Katz2020,Flury2022}, \citet{Flury2022} found a high fraction ($>50\%$) of ionizing-photon leakers in highest-O32 sample (O32 $>10$). 
Further, \citet{Hu2023} presented an anti-correlation between Ly$\alpha$ peak separation and O32 ratio \citep[see also][]{Jaskot2019}. 
Although the correlation might be disturbed by the orientation of density-bounded halos, \citet{Hu2023} found that the highest-O32 sample (O32 $>10$) has the smallest Ly$\alpha$ peak separation $<250$ \kms, supporting the presence of density-bounded channels.
In this work, we obtain an O32 ratio of $9.84\pm 0.79$ from the composite spectrum, close to the criterion suggested by previous works.  This would imply that \textit{on average} the galaxies in our sample have an O32 ratio indicative of non-zero LyC escape. 

The resonance line \civ\ has recently been proposed as a promising probe because the strong \civ\ line emission is detected in six out of eight ionizing-photon leakers \citep[][see also \citealp{Saxena2022,Mascia2023}]{Schaerer2022}. This is for several proposed reasons.  First, the observation of \civ\ might suggest the absence of high-column density gas which could scatter the \civ\ photons.
Second, the \civ/\ciii\ (C43) ratio probes the ionization structure, similar to the O32 ratio.
In this work, we observe a C43 ratio of $0.874\pm0.296$, marginally larger than the criterion (C43 $>0.75$) suggested by \citet{Schaerer2022}.  This again suggests that \textit{on average} the galaxies in our sample have the conditions for LyC escape observed in other local and moderate-redshift galaxy samples, 

Therefore, combining the above evidence, we conclude that indeed there is a moderate fraction of galaxies in our sample that should be the ionizing photon leakers.  At the moment our best constraints come from the O32 and C43 line ratios are just at or slightly above the critical values proposed in previous works.  This evidence is indirect as we must compare the properties of our galaxies to those from lower redshift studies.  Moreover, because we used a stacked spectrum, the exact fraction of galaxies that are leakers, and the fraction of LyC radiation that escapes from them, are unknown.  
Assuming an extreme case, because we adopt the median stacking, we would expect $\sim50\%$ of galaxies having O32 $>10$ and $C43>0.75$.
Considering that the fraction of ionizing-photon leakers in local galaxies with O32 $>10$ is $\sim50\%$ \citep{Flury2022}, we would expect that $\sim25\%$ galaxies in our sample are ionizing-photon leakers.
Recently, \citet{Mascia2023b} also studied the ionizing photon escape of CEERS galaxies at $5.6<z<9$ based on an empirical relation between $f^\mathrm{esc}_\mathrm{LyC}$ and UV slope, \hb\ equivalent width, and galaxy size. They found that only $\sim22\%$ galaxies have $f^\mathrm{esc}_\mathrm{LyC}>20\%$, consistent with our conclusion.  However, much work needs to be done in this area to determine the LyC leakage from individual galaxies in the epoch of reionization.

\section{Summary}

In this work, we construct a composite spectrum based on the JWST CEERS and JADES NIRSpec M-Grating spectra of 63 galaxies at $5.6<z<9$ with a median redshift of $z_\mathrm{med}=6.33$. 
The composite spectrum covers from rest-frame 1500 \AA\ to 5200 \AA, and reliably detects high-ionization UV emission lines, such as \civ, \heii, \oiii, \ciii, and strong optical emission lines, such as \neiii, \oii, \foiii, \hb.
Those emission lines enable us to study the average ISM properties of high-redshift galaxies, such as the chemical abundance, ionization status, and ionizing photon escape. 
The major results are listed as follows:

\begin{itemize}
    \item For the nebular gas, we derive an average dust attenuation $E(B-V)_\mathrm{gas}=0.16^{+0.10}_{-0.11}$ from \hb/\hg, an average electron density $n_e = 570^{+510}_{-290}$ cm$^{-3}$ from the \oii\ doublet ratio, an electron temperature $T_e = 17000^{+1500}_{-1500}$ K from the\foiii~$\lambda4363$/\foiii~$\lambda5007$ ratio, and an ionization parameter $\log(U)=-2.18^{+0.03}_{-0.03}$ from the \foiii\ $\lambda\lambda4959,5007$/\oii\ $\lambda\lambda3727,3729$ ratio. 
    
    \item Using a direct method to determine the electron temperature with the ISM conditions and dust attenuation derived in this work, we calculate an oxygen abundance $12+\log\mathrm{(O/H)}=7.672\pm0.083$ from the \oiii\ $\lambda\lambda4939,5007$/\hb\ and \oii\ $\lambda\lambda3727,3729$/\hb ratios. We also calculate a carbon-to-oxygen (C/O) abundance ratio $\log\mathrm{(C/O)}=-0.87^{+0.13}_{-0.10}$ from the ratio of \ciii\ $\lambda\lambda1907,1909$/\oiii\ $\lambda1666$.
    However, the ICF derived by the \civ\ $\lambda\lambda1548,1551$/\ciii\ $\lambda\lambda1907,1909$ is larger than that derived by the \foiii\ $\lambda\lambda4959,5007$/\oii\ $\lambda\lambda3727,3729$. This might suggest that as much as 91 -- 95\% of the observed \civ\ emission-line flux originates from stellar photospheres and not the nebular gas.
    
    \item 
    Compared to the lower-redshift star-forming galaxies, the composite spectrum of galaxies at $z_\mathrm{med} = 6.33$ here indicates a smaller C/O ratio at fixed oxygen abundance, albeit with moderate significance.
    Considering that the oxygen is released by the supernova of massive stars with lifetimes of  $\sim10$ Myr while the carbon can also be released by the longer-lived AGB stars with lifetimes of  $>200$ Myr, this result suggests that the stellar mass is assembled quickly, with $\lesssim 100$~Myr, prior to substantial enrichment from the latter. 

    \item We study the optical line-ratio diagnostic diagram of  \foiii\ $\lambda\lambda4959,5007$/\hb\ versus \neiii\ $\lambda3869$/\oii\ $\lambda\lambda3727,3729$, and the UV line-ratio diagnostic diagram of, \ciii~$\lambda\lambda1907,1909$/\heii~$\lambda1640$ versus \oiii\ $\lambda1666$/\heii\ $\lambda1640$.
    Our optical diagnostic results are consistent with previous work that high-redshift galaxies have a much harder ionizing spectrum than the $z\sim0$ -- 3 typical galaxies.  
    The UV line-ratio diagnostic diagram also classifies our composite spectrum as high ionization, with a possible presence of weak AGN.
    The position of the composite spectrum on the UV line-ratio diagnostic diagram overlaps the regions spanned by the local extreme dwarf galaxies, I Zw 18 and SBS 0335-052, and $z\sim2$ -- 4 \heii\ emitters, suggesting high ionization mechanisms.
    It may be possible to study these lower redshift extreme galaxies as analogs of ``normal'' star-forming high-redshift galaxies,  This can provide clues to the ionization mechanisms in high-redshift galaxies.

    \item We use the \foiii\ $\lambda\lambda4959,5007$/\oii\ $\lambda\lambda3727,3729$ (O32) and \civ\ $\lambda\lambda1548,1551$/\ciii\ $\lambda\lambda1907,1909$ (C43) ratios to infer indirectly an estimate of the average ionizing photon leakage of our sample.  These emission-line ratios have been shown to correlate with the LyC radiation escape fraction in low-redshift galaxy studies. 
    The O32 and C43 ratios of the composite spectrum are close to, or marginally larger than, the critical values proposed to indicate LyC radiation leakage. 
    This suggests that at least \textit{some} of the high-redshift galaxies in our sample should be the ionizing-photon leakers. However, it is unlikely they dominate the galaxy population in our sample given that the values are just at the critical values.  Future studies that have direct detections of these line ratios will better constrain the fraction of LyC leakers and the fraction of the LyC radiation that escapes from these galaxies.
\end{itemize}

\begin{acknowledgments}
We wish to thank all our colleagues in the CEERS collaboration for their hard work and valuable contributions on this project.  CP thanks Marsha and Ralph Schilling for generous support of this research.   Portions of this research were conducted with the advanced computing resources provided by Texas A\&M High Performance Research Computing (HPRC, \url{http://hprc.tamu.edu}).  This work benefited from support from the George P. and Cynthia Woods Mitchell Institute for Fundamental Physics and Astronomy at Texas A\&M University.    This work acknowledges support from the NASA/ESA/CSA James Webb Space Telescope through the Space Telescope Science Institute, which is operated by the Association of Universities for Research in Astronomy, Incorporated, under NASA contract NAS5-03127. Support for program No. JWST-ERS01345 was provided through a grant from the STScI under NASA contract NAS5-03127.
\end{acknowledgments}

\vspace{5mm}
\facilities{JWST(NIRSpec)}

\software{astropy \citep{2013A&A...558A..33A,2018AJ....156..123A}, }

\appendix
\section{Object Properties}
Table \ref{tab:4} lists the properties of galaxies used in this work.

\startlongtable
\begin{deluxetable*}{c c c c c c}
    \tablecaption{Properties of Galaxies Used in This Work \label{tab:4}}
    \tablehead{
    \colhead{MSA ID} & \colhead{R.A.} & \colhead{Dec.} & \colhead{$z_\mathrm{spec}$} & \colhead{Mag.} & \colhead{log\ $M_\star/M_\odot$}
    }
    \startdata
    \multicolumn{6}{c}{CEERS} \\
    \hline
24 & 214.897231 & 52.843854 & 8.999 & 28.20$^a$ & $8.60^{+0.18}_{-0.28}$\\
23 & 214.901253 & 52.846996 & 8.880 & 28.45$^a$ & $8.72^{+0.17}_{-0.20}$\\
1025 & 214.967532 & 52.932953 & 8.714 & 26.08$^a$ & $8.81^{+0.15}_{-0.16}$\\
1019 & 215.035392 & 52.890667 & 8.678 & 25.09$^a$ & $9.20^{+0.13}_{-0.13}$\\
1029 & 215.218762 & 53.069862 & 8.610 & 25.72$^b$ & $9.84^{+0.28}_{-0.36}$\\
1149 & 215.089714 & 52.966183 & 8.175 & 26.75$^b$ & $9.81^{+0.34}_{-0.46}$\\
4 & 215.005366 & 52.996697 & 7.993 & 27.72$^a$ & $9.55^{+0.22}_{-0.22}$\\
1027 & 214.882996 & 52.840417 & 7.819 & 26.40$^a$ & $8.69^{+0.19}_{-0.20}$\\
1023 & 215.188413 & 53.033647 & 7.776 & 26.23$^b$ & $10.30^{+0.21}_{-0.28}$\\
689 & 214.999053 & 52.941977 & 7.546 & 25.98$^b$ & $9.82^{+0.30}_{-0.42}$\\
698 & 215.050317 & 53.007441 & 7.471 & 25.28$^b$ & $10.04^{+0.20}_{-0.25}$\\
1163 & 214.990468 & 52.971990 & 7.448 & 26.48$^b$ & $9.18^{+0.33}_{-0.31}$\\
1038 & 215.039717 & 52.901598 & 7.194 & 27.53$^a$ & $8.03^{+0.24}_{-0.26}$\\
499 & 214.813006 & 52.834167 & 7.169 & 29.24$^a$ & $7.71^{+0.27}_{-0.25}$\\
407 & 214.839318 & 52.882566 & 7.029 & 28.30$^a$ & $7.58^{+0.45}_{-0.38}$\\
717 & 215.081406 & 52.972180 & 6.932 & 25.45$^b$ & $9.77^{+0.16}_{-0.21}$\\
1143 & 215.077006 & 52.969504 & 6.927 & 27.05$^b$ & $9.20^{+0.33}_{-0.39}$\\
1064 & 215.177167 & 53.048975 & 6.790 & 27.15$^b$ & $9.03^{+0.40}_{-0.39}$\\
613 & 214.882081 & 52.844349 & 6.729 & 27.33$^a$ & $8.51^{+0.27}_{-0.20}$\\
1414 & 215.128020 & 52.984952 & 6.676 & 25.90$^a$ & $8.92^{+0.21}_{-0.17}$\\
386 & 214.832186 & 52.885082 & 6.614 & 28.31$^a$ & $8.12^{+0.22}_{-0.27}$\\
496 & 214.864737 & 52.871719 & 6.569 & 27.51$^a$ & $8.12^{+0.17}_{-0.18}$\\
1160 & 214.805047 & 52.845877 & 6.568 & 27.07$^b$ & $9.10^{+0.43}_{-0.46}$\\
1115 & 215.162817 & 53.073097 & 6.300 & 27.32$^b$ & $9.17^{+0.44}_{-0.40}$ \\
792 & 214.871768 & 52.833167 & 6.257 & 27.67$^a$ & $8.50^{+0.26}_{-0.26}$\\
67 & 215.015598 & 53.011857 & 6.203 & 28.56$^a$ & $8.36^{+0.18}_{-0.21}$\\
1561 & 215.166097 & 53.070755 & 6.196 & 27.16$^b$ & $9.06^{+0.36}_{-0.36}$\\
428 & 214.824554 & 52.868856 & 6.102 & 27.64$^a$ & $7.60^{+0.26}_{-0.10}$\\
355 & 214.806485 & 52.878826 & 6.099 & 27.04$^a$ & $8.42^{+0.20}_{-0.20}$\\
603 & 214.867249 & 52.836736 & 6.057 & 26.49$^a$ & $8.90^{+0.13}_{-0.17}$\\
648 & 214.899825 & 52.847646 & 6.053 & 28.51$^a$ & $8.40^{+0.17}_{-0.23}$\\
618 & 214.876471 & 52.839412 & 6.050 & 27.13$^a$ & $8.32^{+0.14}_{-0.18}$\\
397 & 214.836183 & 52.882678 & 6.001 & 25.43$^a$ & $8.84^{+0.16}_{-0.13}$\\
676 & 214.908480 & 52.845090 & 5.990 & 28.86$^a$ & $8.05^{+0.13}_{-0.23}$\\
1677 & 215.188738 & 53.064378 & 5.867 & 26.03$^b$ & $9.71^{+0.18}_{-0.25}$\\
403 & 214.828970 & 52.875701 & 5.761 & 26.18$^a$ & $9.60^{+0.10}_{-0.16}$\\
323 & 214.872559 & 52.875948 & 5.666 & 27.98$^a$ & $7.62^{+0.18}_{-0.15}$\\
672 & 214.889680 & 52.832976 & 5.666 & 28.99$^a$ & $9.10^{+0.17}_{-0.17}$\\
515 & 214.878536 & 52.874142 & 5.664 & 28.26$^a$ & $8.37^{+0.17}_{-0.20}$\\
2168 & 215.152602 & 53.057062 & 5.654 & 25.23$^b$ & $10.09^{+0.17}_{-0.17}$\\
513 & 214.819364 & 52.832533 & 5.646 & 28.96$^a$ & $8.34^{+0.14}_{-0.16}$\\
746 & 214.809145 & 52.868483 & 5.623 & 29.36$^a$ & $10.25^{+0.23}_{-0.19}$\\
\hline
\multicolumn{6}{c}{JADES}\\
\hline
21842 & 53.156827 & $-27.767162$ & 7.980 & 28.17$^c$ & $8.37^{+0.17}_{-0.21}$\\
10013682 & 53.167449 & $-27.772034$ & 7.275 & 30.30$^c$ & $7.75^{+0.21}_{-0.25}$\\
10013905 & 53.118327 & $-27.769010$ & 7.197 & 26.77$^d$ & $9.58^{+0.34}_{-0.41}$\\
20961 & 53.134230 & $-27.768916$ & 7.045 & 28.10$^c$ & $7.75^{+0.19}_{-0.17}$\\
4297 & 53.155794 & $-27.815209$ & 6.714 & 28.52$^c$ & $7.83^{+0.20}_{-0.22}$\\
3334 & 53.151381 & $-27.819165$ & 6.706 & 28.58$^c$ & $8.31^{+0.09}_{-0.13}$\\
16625 & 53.169047 & $-27.778834$ & 6.631 & 28.26$^c$ & $7.66^{+0.19}_{-0.20}$\\
18846 & 53.134918 & $-27.772711$ & 6.335 & 26.90$^c$ & $7.72^{+0.16}_{-0.11}$\\
18976 & 53.166602 & $-27.772402$ & 6.327 & 28.10$^c$ & $7.58^{+0.22}_{-0.19}$\\
17566 & 53.156101 & $-27.775881$ & 6.102 & 26.64$^c$ & $9.60^{+0.09}_{-0.10}$\\
19342 & 53.160623 & $-27.771611$ & 5.974 & 28.02$^c$ & $7.75^{+0.18}_{-0.16}$\\
10013618 & 53.119112 & $-27.760802$ & 5.944 & 26.67$^d$ & $9.62^{+0.34}_{-0.29}$\\
6002 & 53.110417 & $-27.808924$ & 5.937 & 27.93$^c$ & $7.68^{+0.16}_{-0.15}$\\
9422 & 53.121757 & $-27.797638$ & 5.936 & 27.09$^c$ & $7.61^{+0.10}_{-0.09}$\\
10013704 & 53.126538 & $-27.818090$ & 5.920 & 27.83$^c$ & $8.85^{+0.14}_{-0.16}$\\
10013620 & 53.122590 & $-27.760569$ & 5.917 & 26.58$^d$ & $9.61^{+0.22}_{-0.24}$\\
19606 & 53.176568 & $-27.771131$ & 5.889 & 28.39$^c$ & $7.78^{+0.17}_{-0.16}$\\
10005113 & 53.167302 & $-27.802874$ & 5.821 & 28.76$^c$ & $7.28^{+0.21}_{-0.14}$\\
10056849 & 53.113511 & $-27.772836$ & 5.814 & 27.64$^d$ & $9.15^{+0.33}_{-0.36}$\\
22251 & 53.154072 & $-27.766072$ & 5.798 & 27.63$^d$ & $8.16^{+0.14}_{-0.14}$\\
4404 & 53.115379 & $-27.814774$ & 5.764 & 27.40$^c$ & $7.66^{+0.14}_{-0.16}$\\
    \hline
    \enddata
    \tablecomments{
    \begin{itemize}
        \item[$^a$] JWST NIRCam F150W magnitude from the CEERS survey \citep{Finkelstein2023}.
        \item[$^b$] HST WFC3 F160W magnitude from \citet{Finkelstein2022}
        \item[$^c$] JWST NIRCam F150W magnitude from the JADES survey data release \citep{Eisenstein2023}.
        \item[$^d$] HST WFC3 F160W magnitude from \citet{Whitaker2019}.
    \end{itemize}}
\end{deluxetable*}

\bibliography{main}{}
\bibliographystyle{aasjournal}

\end{document}